\documentclass[reprint,superscriptaddress,amsmath,amssymb,aps,prl]{revtex4-2}

\usepackage{graphicx}
\usepackage{dcolumn}
\usepackage{bm}
\usepackage[colorlinks]{hyperref}
\newcommand{\matr}[1]{\bm{#1}}
\newcommand{\ket}[1]{| #1 \rangle}
\newcommand{\bra}[1]{\langle #1 |}
\newcommand{\braket}[2]{\langle #1 | #2 \rangle}

\begin{document}

\preprint{APS/123-QED}

\title{Noise Constraints for Nonlinear Exceptional Point Sensing}

\author{Xu Zheng}
\affiliation{%
 Division of Physics and Applied Physics, School of Physical and Mathematical Sciences, Nanyang Technological University, Singapore 637371, Singapore
}
\author{Y.~D.~Chong}%
 \email{yidong@ntu.edu.sg}
\affiliation{%
 Division of Physics and Applied Physics, School of Physical and Mathematical Sciences, Nanyang Technological University, Singapore 637371, Singapore
}%
\affiliation{Centre for Disruptive Photonic Technologies, Nanyang Technological University, Singapore 637371, Singapore}

\date{\today}

\begin{abstract}
  Exceptional points (EPs) are singularities in the parameter space of a non-Hermitian system where eigenenergies and eigenstates coincide. They hold promise for enhancing sensing applications, but this is limited by the divergence of shot noise near EPs.  According to recent studies, EP sensors operating in the nonlinear regime may avoid these limitations.  By analyzing an exemplary nonlinear system, we show that the interplay of noise and nonlinearity introduces previously-unidentified obstacles to enhanced sensing.  The noise effectively displaces the EP in parameter space and reduces its order, thereby eliminating the sought-for divergence in the signal-to-noise ratio.  Moreover, the noise near the nonlinear EP experiences a stronger divergence than predicted by standard calculations of the Petermann noise factor, due to the properties of the Bogoliubov-de Gennes Hamiltonian governing the fluctuations.  Our semi-analytical estimates for the noise level agree quantitatively with the results of stochastic numerical simulations.
\end{abstract}

\maketitle


An exceptional point (EP) is a degeneracy of a non-Hermitian system where two or more eigenvalues, as well as their eigenvectors, coalesce. EPs are tied to many interesting phenomena in classical and quantum physics \cite{dembowski2004encircling, heiss2012physics, doppler2016dynamically, goldzak2018light, el2018non, miri2019exceptional, li2020hamiltonian, ozdemir2019parity, chen2020revealing, parto2020non, wang2021coherent}, including light trapping \cite{goldzak2018light}, chiral mode switching \cite{li2020hamiltonian}, and coherent perfect absorption \cite{wang2021coherent}.  One of their most important potential applications is ``EP sensing'': using eigenenergies near EPs to precisely measure physical perturbations \cite{wiersig2014enhancing, wiersig2016sensors, liu2016metrology, xiao2019enhanced}.  Such eigenenergies vary as a square or higher root with the sensing parameter---much faster than the eigenenergy variations of Hermitian systems, which are linear at best.

Although the rapid variation of eigenenergy splittings near EPs has been observed in experiments \cite{zhang2016parity, chen2017exceptional, hodaei2017enhanced, lai2019observation}, there is an ongoing controversy over whether this enables enhanced sensing. The key issue is the effect of noise.  For passive EP sensors, several authors have argued that diverging noise at the EP eliminates any purported parametric enhancement in sensitivity \cite{langbein2018no, lau2018fundamental, zhang2019quantum, chen2019sensitivity, duggan2022limitations, loughlinExceptionalPointSensorsOffer2024}.  These arguments focus on quantum shot noise, which is present in every system and cannot be eliminated \cite{lau2018fundamental, zhang2019quantum, chen2019sensitivity}. In experiments on ring lasers, this noise-induced degradation of sensitivity near an EP has indeed been found \cite{wang2020petermann}.  It has been suggested that such limitations can be overcome by operating near a lasing transition \cite{zhang2019quantum, chen2019sensitivity} or by incorporating nonreciprocal couplings \cite{lau2018fundamental}, but these proposals have not yet been experimentally tested, and in any case EPs may not even be necessary for them.


Another promising avenue for EP sensing is to use nonlinear systems.  This is motivated by two lines of reasoning.  First, as mentioned above, there are signs that operating near the lasing transition is advantageous \cite{zhang2019quantum, chen2019sensitivity}, and above-threshold lasers are inherently nonlinear \cite{Salt2010}.  Second, the predictions of noise divergences near EPs rely on analyzing linear problems (e.g., Petermann noise amplification factors \cite{petermann1979calculated, haus1985, siegman1989excess1, siegman1989excess2, wang2020petermann, wiersig2023petermann} calculated from eigenmodes of a linearized Hamiltonian), which may be invalidated by nonlinearity \cite{peters2022exceptional}.  Specifically, the instantaneous Hamiltonian may lie away from an EP, or exhibit a weakened divergence near the EP \cite{bai2023nonlinear, bai2023nonlinearity, bai2024observation}.

Here, we analyze an exemplary model of nonlinear EP sensing using two coupled cavities with nonlinear gain and/or loss.  The nonlinear steady-state solutions are governed by a third-order EP, and a standard analysis predicts that the eigenvalue susceptibility diverges faster than the noise, so that the system works as a nonlinear EP sensor \cite{bai2023nonlinear, bai2023nonlinearity, bai2024observation}.  However, we show that when the noise-nonlinearity interactions are treated more carefully, the EP shifts in parameter space by an amount dependent on the model's noise level.  This shift is reproduced in numerical simulations, and it transforms the EP into a second-order EP that does not support a divergent signal-to-noise ratio.  We also find another unexpected phenomenon: the noise near the shifted EP has a stronger scaling---inverse, rather than inverse root---than predicted by the standard noise model.  This happens because, near the nonlinear EP, the Bogoliubov-de Gennes Hamiltonian governing weak fluctuations gives a very different Petermann factor \cite{petermann1979calculated, haus1985} than the conventionally-used linearized Hamiltonian.  After correcting for this, our theory achieves quantitative agreement with the frequency noise found in stochastic time-domain simulations, with no fitting parameters.  These results impose strong constraints on EP sensing in nonlinear systems such as coupled-cavity lasers.

\begin{figure}
    \centering
    \includegraphics[width=\linewidth]{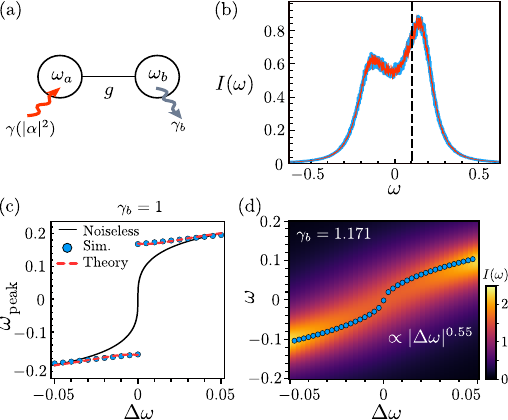}
    \caption{(a) Schematic of an oscillator with nonlinear gain coupled to a linear damped oscillator.  (b) Power spectrum $I(\omega) = \langle|\alpha(\omega)|^2\rangle$ obtained from two separate time-domain simulations at $\Delta\omega=0.005$, $\gamma_b=2g=1$.  Vertical dashes indicate the steady-state frequency predicted by Eq.~\eqref{omegapoly}.  (c)--(d) Peak frequency and power spectrum near (c) the original EP at $\gamma_b=1$ and (d) the shifted EP at $\gamma_b=1.171$. In (c), blue dots denote the frequency of the dominant peak extracted from the simulated spectra via curve-fitting; red dashes show the peak frequencies adjusted by nonlinearity-noise interactions, as described in the text.  The other branch of eigenvalues, whose real parts have opposite signs, is omitted. Solid lines show the stable steady-state frequency of the noise-free model. In (d), the heatmap shows the power spectrum averaged over 100 simulation runs. Blue dots show the dominant peak frequency extracted by curve-fitting. The model parameters are $\gamma_a=2$, $g=0.5$, $\Gamma=0.01$, and $D=1$.}
    \label{fig:spectrumheatmap}
\end{figure}

The model we study is depicted in Fig.~\ref{fig:spectrumheatmap}(a).  It consists of two coupled resonators: a nonlinear resonator with natural frequency $\omega_a$ and nonlinear gain $\gamma$, and a linear one with natural frequency $\omega_b$ and loss rate $\gamma_b$.  It evolves according to the Langevin equations
\begin{align}
  \dot{\alpha} &= \left[-i\omega_a + \frac{1}{2}\gamma(|\alpha|^2)\right]\alpha - ig\beta + \xi_\alpha,
  \label{eq:classicallangevineq1} \\
  \dot{\beta} &= \left(-i\omega_b - \frac{\gamma_b}{2}\right)\beta - ig\alpha + \xi_\beta,
  \label{eq:classicallangevineq2}
\end{align}
where $\alpha,\, \beta$ are the respective $c$-number mode amplitudes for the two oscillators, $g$ is a coupling strength, and $\xi_\alpha,\, \xi_\beta$ are white noise terms that have zero mean and satisfy
\begin{align}
\langle\xi^{\ast}_{\mu}(t)\xi_{\nu}(t')\rangle=D_{\mu\nu}\delta(t-t'),~~~\langle\xi_{\mu}(t)\xi_{\nu}(t')\rangle=0.
\label{eq:classicalnoise}
\end{align}
Here, $D_{\alpha\alpha}$ and $D_{\beta\beta}$ are nonzero, while all other elements give zero. As for the nonlinear gain, our analysis will apply to a variety of functional forms for $\gamma(|\alpha|^2)$; we will present numerical simulations using the gain function of a van der Pol oscillator \cite{holmes1978bifurcations, lee2013quantum, walter2015quantum, dutta2019critical}, but in the Supplemental Materials we show results using laser-type saturable gain, which work equally well with the theory \cite{supplement}.

Applying Eq.~\eqref{eq:classicallangevineq1}--\eqref{eq:classicallangevineq2} to a hypothetical steady-state noise-free solution
($\xi_{\mu} = 0$), the frequency $\omega$ obeys
\begin{equation}
4(\omega-\omega_a)(\omega-\omega_b)^2 + (\omega-\omega_a)\gamma_b^2 - 4(\omega-\omega_b)g^2 = 0.
\label{omegapoly}
\end{equation}
This has a triple root at $\Delta\omega \equiv \omega_a - \omega_b=0$, $\gamma_b=2g$, corresponding to a nonlinear third-order EP \cite{bai2023nonlinear, bai2023nonlinearity,bai2024observation}.  Near the EP, $\omega \approx \omega_b + g|\Delta \omega/g|^{1/3}\operatorname{sgn}(\Delta\omega)$, implying a diverging susceptibility $|\partial\omega / \partial\Delta\omega| \sim \Delta\omega^{-2/3}$ \cite{wiersig2014enhancing}.

To find the noise level in this nonlinear system, we can take the approach of previous studies like Refs.~\cite{bai2023nonlinear, bai2023nonlinearity}.
We introduce the instantaneous Hamiltonian
\begin{align}
  \bm{H} = \left[\begin{array}{cc}
        \omega_a+\frac{i}{2}\gamma_0 & g\\
        g & \omega_b-\frac{i}{2}\gamma_b
    \end{array}\right],
    \label{H0}
\end{align}
where $\gamma_0 = \gamma(|\alpha_0|^2)$ with $\alpha_0$ being the
steady-state field. Near the EP,
\begin{equation}
  \gamma_0 \approx 2g(1-\big|\Delta\omega/g\big|^{2/3}).
  \label{gamma_eff_base}
\end{equation}
The excess noise, caused by eigenstate non-orthogonality, is governed by the Petermann factor \cite{petermann1979calculated, haus1985, siegman1989excess1, siegman1989excess2}
\begin{align}
  K = \frac{1}{|\langle\psi^L|\psi^R\rangle|^2},
  \label{petermannfac}
\end{align}
where $\langle\psi^L|$ and $|\psi^R\rangle$ are respectively the power-normalized left and right eigenstates of $\bm{H}$.  Near the EP, we find $K \sim |\Delta\omega|^{-2/3}$ (see Supplemental Materials \cite{supplement}).  The frequency uncertainty for an oscillator is $\sigma_{\omega} \propto K^{1/2}$ \cite{chow1985ring}, so we have a signal-to-noise ratio
\begin{align}
  \text{SNR} = \frac{|\partial\omega / \partial\Delta\omega|}{\sigma_{\omega}} \sim |\Delta \omega|^{-1/3}.
    \label{SNR}
\end{align}
By contrast, similar analyses for linear systems show $\sigma_\omega$ diverges as rapidly as the susceptibility, leaving the SNR finite \cite{lee2008divergent,langbein2018no, lau2018fundamental,smith2022beyond}.  Note that nonlinearity is not a sufficient condition; if the present model is detuned from the EP by varying $\gamma_b$, the SNR is non-divergent \cite{supplement}.

The above derivation of the noise scaling closely follows the approach of previous studies \cite{lee2008divergent, wang2020petermann, smith2022beyond, wiersig2023petermann, bai2023nonlinear, bai2023nonlinearity, bai2024observation}. However, we find it is \textit{not} accurate near the EP.  We will present numerical results showing discrepancies with the established theory, then discuss how to fix the theory.

We solve the Langevin equations~\eqref{eq:classicallangevineq1}--\eqref{eq:classicallangevineq2} numerically using the Runge-Kutta method with adaptive time steps, via the \texttt{DifferentialEquations.jl} Julia package \cite{rackauckas2017differentialequations}.  For the nonlinear gain, we take the van der Pol form $\gamma=\gamma_a+2\Gamma(1-|\alpha|^2)$ \cite{holmes1978bifurcations,lee2013quantum, walter2015quantum, dutta2019critical},
while the noise is modeled as a Wiener process \cite{rossler2009second}. From each simulation run we derive a Fourier power spectrum (for details, see the Supplemental Materials \cite{supplement}); Fig.~\ref{fig:spectrumheatmap}(b) shows the results from two independent representative runs at $\Delta\omega=0.005$, $\gamma_b=2g$.  We see two asymmetric spectral peaks, with the larger and narrower peak near, but not truly matching, the stable steady-state frequency predicted by Eq.~\eqref{omegapoly} (vertical dashes).  For each choice of parameters, we perform $N = 100$ simulation runs and perform a curve-fit on each of the $N$ spectra to extract the dominant peak frequency.  This yields both a mean peak frequency, and a standard deviation that serves as an \textit{ab initio} estimate for $\sigma_\omega$ (i.e., the ``error bar'' for the location of the peak).  Details about the spectral fitting and peak estimation are given in the Supplemental Materials \cite{supplement}.

We first focus on the mean estimate for the peak.  In Fig.~\ref{fig:spectrumheatmap}(c), these numerically-obtained frequencies are plotted as blue dots, against $\Delta\omega$ for $\gamma_b=2g=1$.  The results do not match the third-order EP behavior given by Eq.~\eqref{omegapoly} (black curve).  This stems from an effect that seems to have been omitted in previous studies of nonlinear EP sensing: the interplay of noise and nonlinearity causes the effective Hamiltonian to deviate substantially from the noise-free instantaneous Hamiltonian \cite{van1994noise}, shifting the EP in parameter space.  To quantify this, let us re-express the mode amplitudes as $\alpha = A\exp(-i\phi_a)$ and $\beta = B\exp(-i\phi_b)$, and convert Eqs.~\eqref{eq:classicallangevineq1}--\eqref{eq:classicallangevineq2} to the form
\begin{align}
  \dot{A}&=\frac{1}{2}\gamma(A^2)A +gB\sin{\varphi}+\xi_A, \label{Adot} \\
  \dot{B}&=-\frac{\gamma_b}{2}B -gA\sin{\varphi}+\xi_B, \label{Bdot} \\
  \dot{\varphi}
  &= \Delta\omega - g\left(\frac{A}{B}-\frac{B}{A}\right) \cos{\varphi}
  +\xi_{\varphi},
  \label{Phidot}
\end{align}
where $\varphi\equiv\phi_a-\phi_b$ is the relative phase, and $\xi_A, \xi_B, \xi_{\varphi}$ are a set of transformed noise variables \cite{supplement}.  Next, we expand $A(t)=A_0+\delta A(t)$, and likewise for $B(t)$ and $\varphi(t)$.  The steady-state values $A_0$, $B_0$, $\varphi_0$ are time-independent quantities that we aim to determine self-consistently, and are \textit{not} assumed to be the same as in the noise-free case.  Eqs.~\eqref{Adot}--\eqref{Phidot} now take on the form
\begin{equation}
  \frac{d}{dt}\begin{pmatrix}\delta A\\\delta B\\\delta \varphi
  \end{pmatrix} = F(A_0,B_0,\varphi_0, \delta A, \delta B, \delta \varphi),
  \label{abphi_eq}
\end{equation}
where $F$ can be expanded up to a desired order in the fluctuations.  The various nonlinear terms are affected by the fluctuations; for example, up to second order,
\begin{align}
  \langle\gamma(A^2)\rangle
  \approx \gamma(A_0^2)
  + \left[\gamma^{\prime}(A_0^2)+2A_0^2\gamma^{\prime\prime}(A_0^2)\right]\langle
  \delta A^2\rangle,
  \label{eq:gammaaeff2}
\end{align}
where $\gamma^{\prime},~\gamma^{\prime\prime}$ are the first and second derivatives of $\gamma$.  If we take $F$ to first order, Eq.~\eqref{abphi_eq} has the form of a generalized Brownian oscillator, and we can solve for the mean squared fluctuations $\langle\delta A^2\rangle$, $\langle\delta\varphi^2\rangle$, and $\langle\delta B^2\rangle$ in terms of $A_0$, $B_0$, and $\varphi_0$.  Conversely, we can expand $F$ to second order and take the time-average of Eq.~\eqref{abphi_eq}; assuming all first-order mean fluctuations ($\langle\delta\dot{A}\rangle$, $\langle\delta A\rangle$, etc.)~vanish, we then obtain $A_0$, $B_0$, and $\varphi_0$ in terms of the mean squared fluctuations.  By combining both procedures, we can estimate the effective gain in Eq.~\eqref{eq:gammaaeff2}, and use this to replace $\gamma_0$ in the Hamiltonian \eqref{H0} to find the adjusted peak frequencies.  For full details about this calculation, see the Supplemental Materials \cite{supplement}.

In Fig.~\ref{fig:spectrumheatmap}(c), the adjusted peak frequencies are plotted as red dashes, and can be seen to agree well with the simulation results (blue dots).  Thus, the absence of an EP for these parameter settings ($\gamma_b = 2g$), contrary to Eq.~\eqref{omegapoly}, can be attributed to the interplay of nonlinearity and noise.  The above theoretical analysis also allows us to estimate where the EP has relocated to: by using \eqref{eq:gammaaeff2} in \eqref{H0}, we find that the EP now occurs when $\Delta \omega = 0$ and
\begin{align}
  \gamma_b \approx 2g \left(1 + \sqrt{\langle\delta\varphi^2\rangle+\frac{\gamma^{\prime}(A_0^2)\, \langle\delta A^2\rangle}{g}}\right),
    \label{eq:gammacritical}
\end{align}
where $\langle\delta A^2\rangle$, $\langle\delta\varphi^2\rangle$, and $A_0^2$ are obtained from the self-consistent calculation described above.

The shifted EP is observed in numerical simulation results, as shown in Fig.~\ref{fig:spectrumheatmap}(d).  Around this EP, the mean peak frequencies obtained from curve-fitting have a scaling of around $|\Delta \omega|^{0.55}$, implying that the EP has been reduced from third-order to second-order.  This is consistent with the noise-adjusted theory, wherein the effective gain \eqref{eq:gammaaeff2}, near the EP \eqref{eq:gammacritical}, takes the form
\begin{equation}
  \langle\gamma\rangle \approx C + \chi\Delta\omega^2,
  \label{chidef}
\end{equation}
unlike the noise-free case \eqref{gamma_eff_base}.  As a result, the leading-order contribution to Eq.~\eqref{H0} becomes $\Delta\omega$ (coming from the detuning term on the diagonal), and the peak frequency scales as $\sim |\Delta\omega|^{1/2}$ \cite{supplement}.

\begin{figure}
  \centering
  \includegraphics[width=\linewidth]{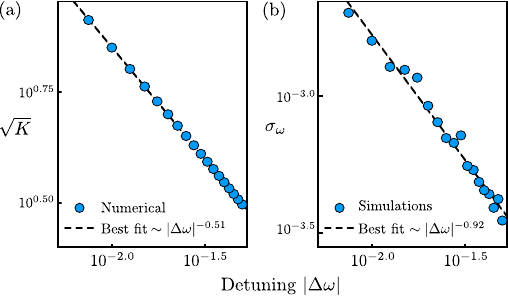}
  \caption{(a) Scaling of the Petermann factor calculated from the instantaneous Hamiltonian~\eqref{H0}.  Here, the shifted EP is taken to be $\gamma_b = 1.182$ to match the analytically-predicted instantaneous Hamiltonian of the effective theory.  (b) Scaling of the frequency uncertainty near the shifted EP, obtained from numerical simulations.  Here we take $\gamma_b = 1.171$, consistent with Fig.~\ref{fig:spectrumheatmap}(d).  Each data point is obtained by taking the standard deviation of 100 simulation runs, as explained in the main text.  In both subplots, dashes show the fitted power laws, and the model parameters are $\gamma_a=2$, $g=0.5$, $\Gamma=0.01$, and $D=1$. }
    \label{fig:noise_scaling}
\end{figure}

We now encounter a second serious discrepancy, involving the frequency noise near the shifted EP.  Calculating the Petermann factor from $\bm{H}$, using the effective gain \eqref{eq:gammaaeff2} matching the shifted EP, we find that the frequency noise should go as $\sigma_\omega \sim K^{1/2} \sim |\Delta\omega|^{-0.51}$, as shown in Fig.~\ref{fig:noise_scaling}(a). This is the noise scaling expected for a linear second-order EP, which is further confirmed by numerical simulations of coupled linear cavities \cite{supplement}.  But the simulation results, shown in Fig.~\ref{fig:noise_scaling}(b), give $\sigma_\omega \sim |\Delta\omega|^{-0.92}$, a faster scaling (i.e., more detrimental to sensing).

The problem apparently stems from the extremely widespread practice of calculating the Petermann factor $K$ from the effective Hamiltonian $\bm{H}$ \cite{petermann1979calculated, haus1985, siegman1989excess1, siegman1989excess2, wang2020petermann, wiersig2023petermann}.  It is understood that $K$ represents the excess noise caused by the non-orthogonality of a set of non-Hermitian eigenmodes \cite{haus1985, wiersig2023petermann}; but which eigenmodes are these?  Near a nonlinear EP, where nonlinear and non-Hermitian effects strongly interact, it appears that $K$ has to be calculated from the Bogoliubov-de Gennes (BdG) Hamiltonian governing the fluctuations, and not $\bm{H}$, which only governs the mean amplitudes.  In fact, we can use the BdG Hamiltonian to obtain not only the right noise scaling, but quantitative predictions of the noise level $\sigma_\omega$ that accurately match simulation results.

To derive the BdG Hamiltonian, we return to Eqs.~\eqref{eq:classicallangevineq1}--\eqref{eq:classicallangevineq2} and let $\alpha=A_0e^{-i\omega_0t}+\delta\alpha$, $\beta=B_0e^{-i\omega_0t+i\varphi_0}+\delta\beta$, where $\omega_0$ is the steady-state frequency.  To leading order, and in a frame rotating at frequency $\omega_0$,
\begin{align}
  \delta\dot{\alpha}&=\left[-i\tilde{\omega}_a+\frac{1}{2}\langle\gamma(A^2)\rangle+\frac{1}{2}A_0^2\gamma^{\prime}(A_0^2)\right]\delta\alpha\nonumber\\
  &~~~~~~~~~~~~+\frac{1}{2}A_0^2\gamma^{\prime}(A_0^2)\delta\alpha^{\ast}-ig\delta\beta+\tilde{\xi}_{\alpha},
  \label{eq:fluctuationdynamics1}\\
  \delta\dot{\beta}&=(-i\tilde{\omega}_b-\frac{\gamma_b}{2})\delta\beta-ig\delta\alpha+\xi_{\beta},
  \label{eq:fluctuationdynamics2}
\end{align}
where $\tilde{\omega}_{a,b}=\omega_{a,b}-\omega_0$ and $\tilde{\xi}_\alpha, \xi_\beta$ are noise terms. Noting that these equations also involve the conjugate mode amplitudes, we define $|\Psi\rangle=[\delta\alpha,\delta\beta,\delta\alpha^{\ast},\delta\beta^{\ast}]^T$ and write
\begin{align}
  i|\dot{\Psi}\rangle=\matr{H}_{B}\,|\Psi\rangle+|\xi\rangle.
  \label{eq:fluctuationdynamics3}
\end{align}
The BdG Hamiltonian $\matr{H}_{B}$ has the form
\begin{align}
  \bm{H}_{B}&=\left[\begin{array}{cc}
    \bm{H}_0+\bm{V}_0 & \bm{V}_0\\
    -\bm{V}_0^{\ast} & -\bm{H}_0^{\ast}-\bm{V}_0^{\ast},\\
  \end{array}\right],
  \label{eq:bdg}\\
  \bm{V}_0&=\left[\begin{array}{cc}
    \frac{i}{2}A_0^2\gamma^{\prime}(A_0^2)& 0\\
  0 & 0
  \end{array}\right],
  \label{eq:v0dv}
\end{align}
where $\bm{H}_0$ is given by Eq.~\eqref{H0} but with $\omega_{a,b}$ replaced by $\tilde{\omega}_{a,b}$ and $\gamma_0$ replaced by $\langle\gamma(A^2)\rangle$.

\begin{figure}
  \centering
  \includegraphics[width=\linewidth]{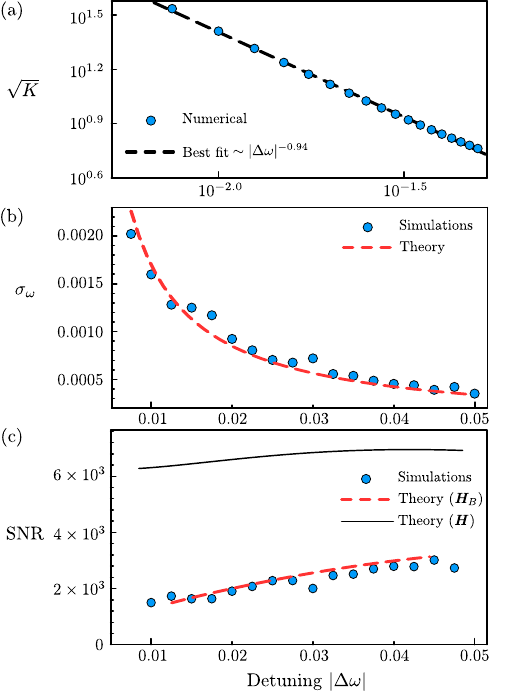}
  \caption{Sensitivity near the shifted EP. (a) $\sqrt{K}$ versus detuning $\Delta\omega$, calculated directly from the BdG Hamiltonian $\matr{H}_{B}$ (blue dots).  The best-fit line scales as $|\Delta\omega|^{-0.94}$ (dashes).  (b) Comparison of frequency uncertainty $\sigma_\omega$ from numerical simulations (blue dots) and the theoretical predication~\eqref{eq:newsigma} (red dashes), using a sampling time of $\Delta t = 10^7$. (c) Comparison of the SNR from numerical simulations (blue dots) and the theoretical prediction \eqref{eq:newsigma} (red dashes).  The SNR predicted from the instantaneous Hamiltonian $\matr{H}$ is much higher (black curve).  None of the SNR's diverge at the shifted EP.  In all subplots, we take $\gamma_b=1.17$ for the simulations, and $\gamma_b=1.18$ for the effective theory, corresponding to the shifted EP under each approach.  All other model parameters are the same as in Fig~\ref{fig:spectrumheatmap}.}
  \label{fig:correctscaling}
\end{figure}

The BdG Hamiltonian has a second order EP, with its position coinciding with the shifted EP. Expressing $\bm{H}_{B}$ to lowest order in $\Delta\omega$ around the EP, we derive the following approximate analytic expression for the Petermann factor:
\begin{align}
  K\approx \frac{-2A_0^2\gamma^{\prime}(A_0^2)g^2}{2g+(\chi g+2)A_0^2\gamma^{\prime}(A_0^2)}\frac{1}{|\Delta\omega|^2}.
  \label{eq:newpeter}
\end{align}
Here $\chi$ is defined in Eq.~\eqref{chidef}, and $A_0$ is the self-consistent mean amplitude.  Hence, the frequency uncertainty should scale as $K^{1/2}\propto |\Delta\omega|^{-1}$.  If we calculate the Petermann factor directly from $\bm{H}_{B}$, we obtain $K^{1/2}\propto|\Delta\omega|^{-0.94}$, as shown in Fig.~\ref{fig:correctscaling}(a).  This is very close to the scaling factor of $-0.92$ obtained from simulations [Fig.~\ref{fig:noise_scaling}(b)].  Going further, we can use Eqs.~\eqref{eq:fluctuationdynamics1}--\eqref{eq:v0dv} to derive the full frequency uncertainty,
\begin{align}
  \sigma_{\omega}\approx\sqrt{\frac{K(D_{\alpha\alpha}+D_{\beta\beta})}{2A_0^2\Delta t}},
  \label{eq:newsigma}
\end{align}
where $\Delta t$ is the measurement time \cite{supplement}.  We can directly compare this to the $\sigma_\omega$ derived from simulations by equating $\Delta t$ to the previously-described sampling time for each simulation run.  As shown in Fig.~\ref{fig:correctscaling}(b), we obtain \textit{quantitative} agreement between theory and simulations, with no additional fitting parameters used in producing the theoretical results.  This cements the claim that $\matr{H}_{B}$, not $\matr{H}$, governs the noise near the nonlinear EP.  Both approaches yield no divergence for the SNR, as shown in Fig.~\ref{fig:correctscaling}(c).  It is also worth noting that the exact variation of the SNR on $|\Delta\omega|$ is dependent on model details, including the noise level parameters $D_{\mu\nu}$.  In the Supplemental Materials, we show that in some cases the SNR may vary non-monotonically with $|\Delta\omega|$, so that there can be a range over which the SNR increases with decreasing $|\Delta\omega|$, as seen in earlier reports \cite{bai2023nonlinearity}.  In all cases, however, we find no actual SNR divergence as $|\Delta\omega| \rightarrow 0$ \cite{supplement}.


In conclusion, we have shown that the operation of nonlinear EP sensors is hampered by the interplay between noise and nonlinearity, which shifts the EP relative to its noise-free position and reduces its order. As a result, there is no enhancement in signal-to-noise ratio as one approaches the EP.  Furthermore, we identify an anomalous and detrimental noise scaling near the nonlinear EP, and trace it to the Bogoliubov-de Gennes Hamiltonian governing fluctuation dynamics.  While we have utilized a specific model with van der Pol nonlinearity, our findings are more broadly applicable. Similar effects are obtained with the laser model of Ref.~\cite{bai2023nonlinearity}, and in a three-cavity model with a fifth-order nonlinear EP \cite{supplement}.  Moreover, even though we have used a classical Langevin framework, similar conclusions are obtained in analogous quantum models \cite{supplement}.   Our theory could be tested experimentally using ring or micropillar optical resonators \cite{supplement}.  Lasing has been well-established in tunable 1D and 2D arrays of such resonators \cite{st-jeanLasingTopologicalEdge2017, partoEdgeModeLasing1D2018, zhaoTopologicalHybridSilicon2018, bandresTopologicalInsulatorLaser2018}; alternatively, nonlinear gain can be introduced through Raman or Brillouin processes \cite{pengLossinducedSuppressionRevival2014, wang2020petermann}. With a better understanding of the limitations of nonlinear EP sensors, it will be possible to explore better noise-mitigation strategies \cite{ramezanpour2021tuning, kononchuk2022exceptional}, other forms of nonlinear singularity-based sensing beyond EPs \cite{peters2022exceptional}, and the combination of quantum sensing techniques with nonlinear EPs \cite{degen2017quantum, pirandola2018advances, lawrie2019quantum, aslam2023quantum, bass2024quantum}.

\begin{acknowledgments}
This work was supported by the Singapore National Research Foundation (NRF) under the NRF Investigatorship NRF-NRFI08-2022-0001, and Competitive Research Program (CRP) Nos.~NRF-CRP23-2019-0005, NRF-CRP23-2019-0007, and NRF-CRP29-2022-0003.
\end{acknowledgments}

\bibliography{reference}

\begin{thebibliography}{64}%
\makeatletter
\providecommand \@ifxundefined [1]{%
 \@ifx{#1\undefined}
}%
\providecommand \@ifnum [1]{%
 \ifnum #1\expandafter \@firstoftwo
 \else \expandafter \@secondoftwo
 \fi
}%
\providecommand \@ifx [1]{%
 \ifx #1\expandafter \@firstoftwo
 \else \expandafter \@secondoftwo
 \fi
}%
\providecommand \natexlab [1]{#1}%
\providecommand \enquote  [1]{``#1''}%
\providecommand \bibnamefont  [1]{#1}%
\providecommand \bibfnamefont [1]{#1}%
\providecommand \citenamefont [1]{#1}%
\providecommand \href@noop [0]{\@secondoftwo}%
\providecommand \href [0]{\begingroup \@sanitize@url \@href}%
\providecommand \@href[1]{\@@startlink{#1}\@@href}%
\providecommand \@@href[1]{\endgroup#1\@@endlink}%
\providecommand \@sanitize@url [0]{\catcode `\\12\catcode `\$12\catcode `\&12\catcode `\#12\catcode `\^12\catcode `\_12\catcode `\%12\relax}%
\providecommand \@@startlink[1]{}%
\providecommand \@@endlink[0]{}%
\providecommand \url  [0]{\begingroup\@sanitize@url \@url }%
\providecommand \@url [1]{\endgroup\@href {#1}{\urlprefix }}%
\providecommand \urlprefix  [0]{URL }%
\providecommand \Eprint [0]{\href }%
\providecommand \doibase [0]{https://doi.org/}%
\providecommand \selectlanguage [0]{\@gobble}%
\providecommand \bibinfo  [0]{\@secondoftwo}%
\providecommand \bibfield  [0]{\@secondoftwo}%
\providecommand \translation [1]{[#1]}%
\providecommand \BibitemOpen [0]{}%
\providecommand \bibitemStop [0]{}%
\providecommand \bibitemNoStop [0]{.\EOS\space}%
\providecommand \EOS [0]{\spacefactor3000\relax}%
\providecommand \BibitemShut  [1]{\csname bibitem#1\endcsname}%
\let\auto@bib@innerbib\@empty
\bibitem [{\citenamefont {Dembowski}\ \emph {et~al.}(2004)\citenamefont {Dembowski}, \citenamefont {Dietz}, \citenamefont {Gr{\"a}f}, \citenamefont {Harney}, \citenamefont {Heine}, \citenamefont {Heiss},\ and\ \citenamefont {Richter}}]{dembowski2004encircling}%
  \BibitemOpen
  \bibfield  {author} {\bibinfo {author} {\bibfnamefont {C.}~\bibnamefont {Dembowski}}, \bibinfo {author} {\bibfnamefont {B.}~\bibnamefont {Dietz}}, \bibinfo {author} {\bibfnamefont {H.-D.}\ \bibnamefont {Gr{\"a}f}}, \bibinfo {author} {\bibfnamefont {H.}~\bibnamefont {Harney}}, \bibinfo {author} {\bibfnamefont {A.}~\bibnamefont {Heine}}, \bibinfo {author} {\bibfnamefont {W.}~\bibnamefont {Heiss}},\ and\ \bibinfo {author} {\bibfnamefont {A.}~\bibnamefont {Richter}},\ }\bibfield  {title} {\bibinfo {title} {{Encircling an exceptional point}},\ }\href@noop {} {\bibfield  {journal} {\bibinfo  {journal} {Phys. Rev. E}\ }\textbf {\bibinfo {volume} {69}},\ \bibinfo {pages} {056216} (\bibinfo {year} {2004})}\BibitemShut {NoStop}%
\bibitem [{\citenamefont {Heiss}(2012)}]{heiss2012physics}%
  \BibitemOpen
  \bibfield  {author} {\bibinfo {author} {\bibfnamefont {W.}~\bibnamefont {Heiss}},\ }\bibfield  {title} {\bibinfo {title} {{The physics of exceptional points}},\ }\href@noop {} {\bibfield  {journal} {\bibinfo  {journal} {J. Phys. A}\ }\textbf {\bibinfo {volume} {45}},\ \bibinfo {pages} {444016} (\bibinfo {year} {2012})}\BibitemShut {NoStop}%
\bibitem [{\citenamefont {Doppler}\ \emph {et~al.}(2016)\citenamefont {Doppler}, \citenamefont {Mailybaev}, \citenamefont {B{\"o}hm}, \citenamefont {Kuhl}, \citenamefont {Girschik}, \citenamefont {Libisch}, \citenamefont {Milburn}, \citenamefont {Rabl}, \citenamefont {Moiseyev},\ and\ \citenamefont {Rotter}}]{doppler2016dynamically}%
  \BibitemOpen
  \bibfield  {author} {\bibinfo {author} {\bibfnamefont {J.}~\bibnamefont {Doppler}}, \bibinfo {author} {\bibfnamefont {A.~A.}\ \bibnamefont {Mailybaev}}, \bibinfo {author} {\bibfnamefont {J.}~\bibnamefont {B{\"o}hm}}, \bibinfo {author} {\bibfnamefont {U.}~\bibnamefont {Kuhl}}, \bibinfo {author} {\bibfnamefont {A.}~\bibnamefont {Girschik}}, \bibinfo {author} {\bibfnamefont {F.}~\bibnamefont {Libisch}}, \bibinfo {author} {\bibfnamefont {T.~J.}\ \bibnamefont {Milburn}}, \bibinfo {author} {\bibfnamefont {P.}~\bibnamefont {Rabl}}, \bibinfo {author} {\bibfnamefont {N.}~\bibnamefont {Moiseyev}},\ and\ \bibinfo {author} {\bibfnamefont {S.}~\bibnamefont {Rotter}},\ }\bibfield  {title} {\bibinfo {title} {{Dynamically encircling an exceptional point for asymmetric mode switching}},\ }\href@noop {} {\bibfield  {journal} {\bibinfo  {journal} {Nature}\ }\textbf {\bibinfo {volume} {537}},\ \bibinfo {pages} {76} (\bibinfo {year} {2016})}\BibitemShut {NoStop}%
\bibitem [{\citenamefont {Goldzak}\ \emph {et~al.}(2018)\citenamefont {Goldzak}, \citenamefont {Mailybaev},\ and\ \citenamefont {Moiseyev}}]{goldzak2018light}%
  \BibitemOpen
  \bibfield  {author} {\bibinfo {author} {\bibfnamefont {T.}~\bibnamefont {Goldzak}}, \bibinfo {author} {\bibfnamefont {A.~A.}\ \bibnamefont {Mailybaev}},\ and\ \bibinfo {author} {\bibfnamefont {N.}~\bibnamefont {Moiseyev}},\ }\bibfield  {title} {\bibinfo {title} {{Light Stops at Exceptional Points}},\ }\href@noop {} {\bibfield  {journal} {\bibinfo  {journal} {Phys. Rev. Lett.}\ }\textbf {\bibinfo {volume} {120}},\ \bibinfo {pages} {013901} (\bibinfo {year} {2018})}\BibitemShut {NoStop}%
\bibitem [{\citenamefont {El-Ganainy}\ \emph {et~al.}(2018)\citenamefont {El-Ganainy}, \citenamefont {Makris}, \citenamefont {Khajavikhan}, \citenamefont {Musslimani}, \citenamefont {Rotter},\ and\ \citenamefont {Christodoulides}}]{el2018non}%
  \BibitemOpen
  \bibfield  {author} {\bibinfo {author} {\bibfnamefont {R.}~\bibnamefont {El-Ganainy}}, \bibinfo {author} {\bibfnamefont {K.~G.}\ \bibnamefont {Makris}}, \bibinfo {author} {\bibfnamefont {M.}~\bibnamefont {Khajavikhan}}, \bibinfo {author} {\bibfnamefont {Z.~H.}\ \bibnamefont {Musslimani}}, \bibinfo {author} {\bibfnamefont {S.}~\bibnamefont {Rotter}},\ and\ \bibinfo {author} {\bibfnamefont {D.~N.}\ \bibnamefont {Christodoulides}},\ }\bibfield  {title} {\bibinfo {title} {{Non-Hermitian physics and PT symmetry}},\ }\href@noop {} {\bibfield  {journal} {\bibinfo  {journal} {Nat. Phys.}\ }\textbf {\bibinfo {volume} {14}},\ \bibinfo {pages} {11} (\bibinfo {year} {2018})}\BibitemShut {NoStop}%
\bibitem [{\citenamefont {Miri}\ and\ \citenamefont {Alu}(2019)}]{miri2019exceptional}%
  \BibitemOpen
  \bibfield  {author} {\bibinfo {author} {\bibfnamefont {M.-A.}\ \bibnamefont {Miri}}\ and\ \bibinfo {author} {\bibfnamefont {A.}~\bibnamefont {Alu}},\ }\bibfield  {title} {\bibinfo {title} {{Exceptional points in optics and photonics}},\ }\href@noop {} {\bibfield  {journal} {\bibinfo  {journal} {Science}\ }\textbf {\bibinfo {volume} {363}},\ \bibinfo {pages} {eaar7709} (\bibinfo {year} {2019})}\BibitemShut {NoStop}%
\bibitem [{\citenamefont {Li}\ \emph {et~al.}(2020)\citenamefont {Li}, \citenamefont {Dong}, \citenamefont {Wang}, \citenamefont {Cheng}, \citenamefont {Ho}, \citenamefont {Zhang}, \citenamefont {Wen}, \citenamefont {Zhang}, \citenamefont {Chan}, \citenamefont {Al{\`u}} \emph {et~al.}}]{li2020hamiltonian}%
  \BibitemOpen
  \bibfield  {author} {\bibinfo {author} {\bibfnamefont {A.}~\bibnamefont {Li}}, \bibinfo {author} {\bibfnamefont {J.}~\bibnamefont {Dong}}, \bibinfo {author} {\bibfnamefont {J.}~\bibnamefont {Wang}}, \bibinfo {author} {\bibfnamefont {Z.}~\bibnamefont {Cheng}}, \bibinfo {author} {\bibfnamefont {J.~S.}\ \bibnamefont {Ho}}, \bibinfo {author} {\bibfnamefont {D.}~\bibnamefont {Zhang}}, \bibinfo {author} {\bibfnamefont {J.}~\bibnamefont {Wen}}, \bibinfo {author} {\bibfnamefont {X.-L.}\ \bibnamefont {Zhang}}, \bibinfo {author} {\bibfnamefont {C.~T.}\ \bibnamefont {Chan}}, \bibinfo {author} {\bibfnamefont {A.}~\bibnamefont {Al{\`u}}}, \emph {et~al.},\ }\bibfield  {title} {\bibinfo {title} {{Hamiltonian Hopping for Efficient Chiral Mode Switching in Encircling Exceptional Points}},\ }\href@noop {} {\bibfield  {journal} {\bibinfo  {journal} {Phys. Rev. Lett.}\ }\textbf {\bibinfo {volume} {125}},\ \bibinfo {pages} {187403} (\bibinfo {year} {2020})}\BibitemShut {NoStop}%
\bibitem [{\citenamefont {{\"O}zdemir}\ \emph {et~al.}(2019)\citenamefont {{\"O}zdemir}, \citenamefont {Rotter}, \citenamefont {Nori},\ and\ \citenamefont {Yang}}]{ozdemir2019parity}%
  \BibitemOpen
  \bibfield  {author} {\bibinfo {author} {\bibfnamefont {{\c{S}}.~K.}\ \bibnamefont {{\"O}zdemir}}, \bibinfo {author} {\bibfnamefont {S.}~\bibnamefont {Rotter}}, \bibinfo {author} {\bibfnamefont {F.}~\bibnamefont {Nori}},\ and\ \bibinfo {author} {\bibfnamefont {L.}~\bibnamefont {Yang}},\ }\bibfield  {title} {\bibinfo {title} {{Parity--time symmetry and exceptional points in photonics}},\ }\href@noop {} {\bibfield  {journal} {\bibinfo  {journal} {Nat. Mater.}\ }\textbf {\bibinfo {volume} {18}},\ \bibinfo {pages} {783} (\bibinfo {year} {2019})}\BibitemShut {NoStop}%
\bibitem [{\citenamefont {Chen}\ \emph {et~al.}(2020)\citenamefont {Chen}, \citenamefont {Liu}, \citenamefont {Luan}, \citenamefont {Liu}, \citenamefont {Wang}, \citenamefont {Zhu}, \citenamefont {Li}, \citenamefont {Gu}, \citenamefont {Liang}, \citenamefont {Gao} \emph {et~al.}}]{chen2020revealing}%
  \BibitemOpen
  \bibfield  {author} {\bibinfo {author} {\bibfnamefont {H.-Z.}\ \bibnamefont {Chen}}, \bibinfo {author} {\bibfnamefont {T.}~\bibnamefont {Liu}}, \bibinfo {author} {\bibfnamefont {H.-Y.}\ \bibnamefont {Luan}}, \bibinfo {author} {\bibfnamefont {R.-J.}\ \bibnamefont {Liu}}, \bibinfo {author} {\bibfnamefont {X.-Y.}\ \bibnamefont {Wang}}, \bibinfo {author} {\bibfnamefont {X.-F.}\ \bibnamefont {Zhu}}, \bibinfo {author} {\bibfnamefont {Y.-B.}\ \bibnamefont {Li}}, \bibinfo {author} {\bibfnamefont {Z.-M.}\ \bibnamefont {Gu}}, \bibinfo {author} {\bibfnamefont {S.-J.}\ \bibnamefont {Liang}}, \bibinfo {author} {\bibfnamefont {H.}~\bibnamefont {Gao}}, \emph {et~al.},\ }\bibfield  {title} {\bibinfo {title} {{Revealing the missing dimension at an exceptional point}},\ }\href@noop {} {\bibfield  {journal} {\bibinfo  {journal} {Nat. Phys.}\ }\textbf {\bibinfo {volume} {16}},\ \bibinfo {pages} {571} (\bibinfo {year} {2020})}\BibitemShut {NoStop}%
\bibitem [{\citenamefont {Parto}\ \emph {et~al.}(2020)\citenamefont {Parto}, \citenamefont {Liu}, \citenamefont {Bahari}, \citenamefont {Khajavikhan},\ and\ \citenamefont {Christodoulides}}]{parto2020non}%
  \BibitemOpen
  \bibfield  {author} {\bibinfo {author} {\bibfnamefont {M.}~\bibnamefont {Parto}}, \bibinfo {author} {\bibfnamefont {Y.~G.}\ \bibnamefont {Liu}}, \bibinfo {author} {\bibfnamefont {B.}~\bibnamefont {Bahari}}, \bibinfo {author} {\bibfnamefont {M.}~\bibnamefont {Khajavikhan}},\ and\ \bibinfo {author} {\bibfnamefont {D.~N.}\ \bibnamefont {Christodoulides}},\ }\bibfield  {title} {\bibinfo {title} {{Non-Hermitian and Topological Photonics: Optics at an Exceptional Point}},\ }\href@noop {} {\bibfield  {journal} {\bibinfo  {journal} {Nanophotonics}\ }\textbf {\bibinfo {volume} {10}},\ \bibinfo {pages} {403} (\bibinfo {year} {2020})}\BibitemShut {NoStop}%
\bibitem [{\citenamefont {Wang}\ \emph {et~al.}(2021)\citenamefont {Wang}, \citenamefont {Sweeney}, \citenamefont {Stone},\ and\ \citenamefont {Yang}}]{wang2021coherent}%
  \BibitemOpen
  \bibfield  {author} {\bibinfo {author} {\bibfnamefont {C.}~\bibnamefont {Wang}}, \bibinfo {author} {\bibfnamefont {W.~R.}\ \bibnamefont {Sweeney}}, \bibinfo {author} {\bibfnamefont {A.~D.}\ \bibnamefont {Stone}},\ and\ \bibinfo {author} {\bibfnamefont {L.}~\bibnamefont {Yang}},\ }\bibfield  {title} {\bibinfo {title} {{Coherent perfect absorption at an exceptional point}},\ }\href@noop {} {\bibfield  {journal} {\bibinfo  {journal} {Science}\ }\textbf {\bibinfo {volume} {373}},\ \bibinfo {pages} {1261} (\bibinfo {year} {2021})}\BibitemShut {NoStop}%
\bibitem [{\citenamefont {Wiersig}(2014)}]{wiersig2014enhancing}%
  \BibitemOpen
  \bibfield  {author} {\bibinfo {author} {\bibfnamefont {J.}~\bibnamefont {Wiersig}},\ }\bibfield  {title} {\bibinfo {title} {{Enhancing the sensitivity of frequency and energy splitting detection by using exceptional points: application to microcavity sensors for single-particle detection}},\ }\href@noop {} {\bibfield  {journal} {\bibinfo  {journal} {Phys. Rev. Lett.}\ }\textbf {\bibinfo {volume} {112}},\ \bibinfo {pages} {203901} (\bibinfo {year} {2014})}\BibitemShut {NoStop}%
\bibitem [{\citenamefont {Wiersig}(2016)}]{wiersig2016sensors}%
  \BibitemOpen
  \bibfield  {author} {\bibinfo {author} {\bibfnamefont {J.}~\bibnamefont {Wiersig}},\ }\bibfield  {title} {\bibinfo {title} {{Sensors operating at exceptional points: General theory}},\ }\href@noop {} {\bibfield  {journal} {\bibinfo  {journal} {Phys. Rev. A}\ }\textbf {\bibinfo {volume} {93}},\ \bibinfo {pages} {033809} (\bibinfo {year} {2016})}\BibitemShut {NoStop}%
\bibitem [{\citenamefont {Liu}\ \emph {et~al.}(2016)\citenamefont {Liu}, \citenamefont {Zhang}, \citenamefont {{\"O}zdemir}, \citenamefont {Peng}, \citenamefont {Jing}, \citenamefont {L{\"u}}, \citenamefont {Li}, \citenamefont {Yang}, \citenamefont {Nori},\ and\ \citenamefont {Liu}}]{liu2016metrology}%
  \BibitemOpen
  \bibfield  {author} {\bibinfo {author} {\bibfnamefont {Z.-P.}\ \bibnamefont {Liu}}, \bibinfo {author} {\bibfnamefont {J.}~\bibnamefont {Zhang}}, \bibinfo {author} {\bibfnamefont {{\c{S}}.~K.}\ \bibnamefont {{\"O}zdemir}}, \bibinfo {author} {\bibfnamefont {B.}~\bibnamefont {Peng}}, \bibinfo {author} {\bibfnamefont {H.}~\bibnamefont {Jing}}, \bibinfo {author} {\bibfnamefont {X.-Y.}\ \bibnamefont {L{\"u}}}, \bibinfo {author} {\bibfnamefont {C.-W.}\ \bibnamefont {Li}}, \bibinfo {author} {\bibfnamefont {L.}~\bibnamefont {Yang}}, \bibinfo {author} {\bibfnamefont {F.}~\bibnamefont {Nori}},\ and\ \bibinfo {author} {\bibfnamefont {Y.~X.}\ \bibnamefont {Liu}},\ }\bibfield  {title} {\bibinfo {title} {{Metrology with PT-symmetric cavities: enhanced sensitivity near the PT-phase transition}},\ }\href@noop {} {\bibfield  {journal} {\bibinfo  {journal} {Phys. Rev. Lett.}\ }\textbf {\bibinfo {volume} {117}},\ \bibinfo {pages} {110802} (\bibinfo {year} {2016})}\BibitemShut {NoStop}%
\bibitem [{\citenamefont {Xiao}\ \emph {et~al.}(2019)\citenamefont {Xiao}, \citenamefont {Li}, \citenamefont {Kottos},\ and\ \citenamefont {Al{\`u}}}]{xiao2019enhanced}%
  \BibitemOpen
  \bibfield  {author} {\bibinfo {author} {\bibfnamefont {Z.}~\bibnamefont {Xiao}}, \bibinfo {author} {\bibfnamefont {H.}~\bibnamefont {Li}}, \bibinfo {author} {\bibfnamefont {T.}~\bibnamefont {Kottos}},\ and\ \bibinfo {author} {\bibfnamefont {A.}~\bibnamefont {Al{\`u}}},\ }\bibfield  {title} {\bibinfo {title} {{Enhanced sensing and nondegraded thermal noise performance based on PT-symmetric electronic circuits with a sixth-order exceptional point}},\ }\href@noop {} {\bibfield  {journal} {\bibinfo  {journal} {Phys. Rev. Lett.}\ }\textbf {\bibinfo {volume} {123}},\ \bibinfo {pages} {213901} (\bibinfo {year} {2019})}\BibitemShut {NoStop}%
\bibitem [{\citenamefont {Zhang}\ \emph {et~al.}(2016)\citenamefont {Zhang}, \citenamefont {Yong}, \citenamefont {Zhang},\ and\ \citenamefont {He}}]{zhang2016parity}%
  \BibitemOpen
  \bibfield  {author} {\bibinfo {author} {\bibfnamefont {S.}~\bibnamefont {Zhang}}, \bibinfo {author} {\bibfnamefont {Z.}~\bibnamefont {Yong}}, \bibinfo {author} {\bibfnamefont {Y.}~\bibnamefont {Zhang}},\ and\ \bibinfo {author} {\bibfnamefont {S.}~\bibnamefont {He}},\ }\bibfield  {title} {\bibinfo {title} {{Parity-time symmetry breaking in coupled nanobeam cavities}},\ }\href@noop {} {\bibfield  {journal} {\bibinfo  {journal} {Sci. Rep.}\ }\textbf {\bibinfo {volume} {6}},\ \bibinfo {pages} {24487} (\bibinfo {year} {2016})}\BibitemShut {NoStop}%
\bibitem [{\citenamefont {Chen}\ \emph {et~al.}(2017)\citenamefont {Chen}, \citenamefont {Kaya~{\"O}zdemir}, \citenamefont {Zhao}, \citenamefont {Wiersig},\ and\ \citenamefont {Yang}}]{chen2017exceptional}%
  \BibitemOpen
  \bibfield  {author} {\bibinfo {author} {\bibfnamefont {W.}~\bibnamefont {Chen}}, \bibinfo {author} {\bibfnamefont {{\c{S}}.}~\bibnamefont {Kaya~{\"O}zdemir}}, \bibinfo {author} {\bibfnamefont {G.}~\bibnamefont {Zhao}}, \bibinfo {author} {\bibfnamefont {J.}~\bibnamefont {Wiersig}},\ and\ \bibinfo {author} {\bibfnamefont {L.}~\bibnamefont {Yang}},\ }\bibfield  {title} {\bibinfo {title} {{Exceptional points enhance sensing in an optical microcavity}},\ }\href@noop {} {\bibfield  {journal} {\bibinfo  {journal} {Nature}\ }\textbf {\bibinfo {volume} {548}},\ \bibinfo {pages} {192} (\bibinfo {year} {2017})}\BibitemShut {NoStop}%
\bibitem [{\citenamefont {Hodaei}\ \emph {et~al.}(2017)\citenamefont {Hodaei}, \citenamefont {Hassan}, \citenamefont {Wittek}, \citenamefont {Garcia-Gracia}, \citenamefont {El-Ganainy}, \citenamefont {Christodoulides},\ and\ \citenamefont {Khajavikhan}}]{hodaei2017enhanced}%
  \BibitemOpen
  \bibfield  {author} {\bibinfo {author} {\bibfnamefont {H.}~\bibnamefont {Hodaei}}, \bibinfo {author} {\bibfnamefont {A.~U.}\ \bibnamefont {Hassan}}, \bibinfo {author} {\bibfnamefont {S.}~\bibnamefont {Wittek}}, \bibinfo {author} {\bibfnamefont {H.}~\bibnamefont {Garcia-Gracia}}, \bibinfo {author} {\bibfnamefont {R.}~\bibnamefont {El-Ganainy}}, \bibinfo {author} {\bibfnamefont {D.~N.}\ \bibnamefont {Christodoulides}},\ and\ \bibinfo {author} {\bibfnamefont {M.}~\bibnamefont {Khajavikhan}},\ }\bibfield  {title} {\bibinfo {title} {{Enhanced sensitivity at higher-order exceptional points}},\ }\href@noop {} {\bibfield  {journal} {\bibinfo  {journal} {Nature}\ }\textbf {\bibinfo {volume} {548}},\ \bibinfo {pages} {187} (\bibinfo {year} {2017})}\BibitemShut {NoStop}%
\bibitem [{\citenamefont {Lai}\ \emph {et~al.}(2019)\citenamefont {Lai}, \citenamefont {Lu}, \citenamefont {Suh}, \citenamefont {Yuan},\ and\ \citenamefont {Vahala}}]{lai2019observation}%
  \BibitemOpen
  \bibfield  {author} {\bibinfo {author} {\bibfnamefont {Y.-H.}\ \bibnamefont {Lai}}, \bibinfo {author} {\bibfnamefont {Y.-K.}\ \bibnamefont {Lu}}, \bibinfo {author} {\bibfnamefont {M.-G.}\ \bibnamefont {Suh}}, \bibinfo {author} {\bibfnamefont {Z.}~\bibnamefont {Yuan}},\ and\ \bibinfo {author} {\bibfnamefont {K.}~\bibnamefont {Vahala}},\ }\bibfield  {title} {\bibinfo {title} {{Observation of the exceptional-point-enhanced Sagnac effect}},\ }\href@noop {} {\bibfield  {journal} {\bibinfo  {journal} {Nature}\ }\textbf {\bibinfo {volume} {576}},\ \bibinfo {pages} {65} (\bibinfo {year} {2019})}\BibitemShut {NoStop}%
\bibitem [{\citenamefont {Langbein}(2018)}]{langbein2018no}%
  \BibitemOpen
  \bibfield  {author} {\bibinfo {author} {\bibfnamefont {W.}~\bibnamefont {Langbein}},\ }\bibfield  {title} {\bibinfo {title} {{No exceptional precision of exceptional-point sensors}},\ }\href@noop {} {\bibfield  {journal} {\bibinfo  {journal} {Phys. Rev. A}\ }\textbf {\bibinfo {volume} {98}},\ \bibinfo {pages} {023805} (\bibinfo {year} {2018})}\BibitemShut {NoStop}%
\bibitem [{\citenamefont {Lau}\ and\ \citenamefont {Clerk}(2018)}]{lau2018fundamental}%
  \BibitemOpen
  \bibfield  {author} {\bibinfo {author} {\bibfnamefont {H.-K.}\ \bibnamefont {Lau}}\ and\ \bibinfo {author} {\bibfnamefont {A.~A.}\ \bibnamefont {Clerk}},\ }\bibfield  {title} {\bibinfo {title} {{Fundamental limits and non-reciprocal approaches in non-Hermitian quantum sensing}},\ }\href@noop {} {\bibfield  {journal} {\bibinfo  {journal} {Nat. Commun.}\ }\textbf {\bibinfo {volume} {9}},\ \bibinfo {pages} {4320} (\bibinfo {year} {2018})}\BibitemShut {NoStop}%
\bibitem [{\citenamefont {Zhang}\ \emph {et~al.}(2019)\citenamefont {Zhang}, \citenamefont {Sweeney}, \citenamefont {Hsu}, \citenamefont {Yang}, \citenamefont {Stone},\ and\ \citenamefont {Jiang}}]{zhang2019quantum}%
  \BibitemOpen
  \bibfield  {author} {\bibinfo {author} {\bibfnamefont {M.}~\bibnamefont {Zhang}}, \bibinfo {author} {\bibfnamefont {W.}~\bibnamefont {Sweeney}}, \bibinfo {author} {\bibfnamefont {C.~W.}\ \bibnamefont {Hsu}}, \bibinfo {author} {\bibfnamefont {L.}~\bibnamefont {Yang}}, \bibinfo {author} {\bibfnamefont {A.~D.}\ \bibnamefont {Stone}},\ and\ \bibinfo {author} {\bibfnamefont {L.}~\bibnamefont {Jiang}},\ }\bibfield  {title} {\bibinfo {title} {{Quantum noise theory of exceptional point amplifying sensors}},\ }\href@noop {} {\bibfield  {journal} {\bibinfo  {journal} {Phys. Rev. Lett.}\ }\textbf {\bibinfo {volume} {123}},\ \bibinfo {pages} {180501} (\bibinfo {year} {2019})}\BibitemShut {NoStop}%
\bibitem [{\citenamefont {Chen}\ \emph {et~al.}(2019)\citenamefont {Chen}, \citenamefont {Jin},\ and\ \citenamefont {Liu}}]{chen2019sensitivity}%
  \BibitemOpen
  \bibfield  {author} {\bibinfo {author} {\bibfnamefont {C.}~\bibnamefont {Chen}}, \bibinfo {author} {\bibfnamefont {L.}~\bibnamefont {Jin}},\ and\ \bibinfo {author} {\bibfnamefont {R.-B.}\ \bibnamefont {Liu}},\ }\bibfield  {title} {\bibinfo {title} {{Sensitivity of parameter estimation near the exceptional point of a non-Hermitian system}},\ }\href@noop {} {\bibfield  {journal} {\bibinfo  {journal} {New J. Phys.}\ }\textbf {\bibinfo {volume} {21}},\ \bibinfo {pages} {083002} (\bibinfo {year} {2019})}\BibitemShut {NoStop}%
\bibitem [{\citenamefont {Duggan}\ \emph {et~al.}(2022)\citenamefont {Duggan}, \citenamefont {Mann},\ and\ \citenamefont {Alù}}]{duggan2022limitations}%
  \BibitemOpen
  \bibfield  {author} {\bibinfo {author} {\bibfnamefont {R.}~\bibnamefont {Duggan}}, \bibinfo {author} {\bibfnamefont {S.~A.}\ \bibnamefont {Mann}},\ and\ \bibinfo {author} {\bibfnamefont {A.}~\bibnamefont {Alù}},\ }\bibfield  {title} {\bibinfo {title} {{Limitations of sensing at an exceptional point}},\ }\href@noop {} {\bibfield  {journal} {\bibinfo  {journal} {ACS Photonics}\ }\textbf {\bibinfo {volume} {9}},\ \bibinfo {pages} {1554} (\bibinfo {year} {2022})}\BibitemShut {NoStop}%
\bibitem [{\citenamefont {Loughlin}\ and\ \citenamefont {Sudhir}(2024)}]{loughlinExceptionalPointSensorsOffer2024}%
  \BibitemOpen
  \bibfield  {author} {\bibinfo {author} {\bibfnamefont {H.}~\bibnamefont {Loughlin}}\ and\ \bibinfo {author} {\bibfnamefont {V.}~\bibnamefont {Sudhir}},\ }\bibfield  {title} {\bibinfo {title} {Exceptional-{{Point Sensors Offer No Fundamental Signal-to-Noise Ratio Enhancement}}},\ }\href@noop {} {\bibfield  {journal} {\bibinfo  {journal} {Phys. Rev. Lett.}\ }\textbf {\bibinfo {volume} {132}},\ \bibinfo {pages} {243601} (\bibinfo {year} {2024})}\BibitemShut {NoStop}%
\bibitem [{\citenamefont {Wang}\ \emph {et~al.}(2020)\citenamefont {Wang}, \citenamefont {Lai}, \citenamefont {Yuan}, \citenamefont {Suh},\ and\ \citenamefont {Vahala}}]{wang2020petermann}%
  \BibitemOpen
  \bibfield  {author} {\bibinfo {author} {\bibfnamefont {H.}~\bibnamefont {Wang}}, \bibinfo {author} {\bibfnamefont {Y.-H.}\ \bibnamefont {Lai}}, \bibinfo {author} {\bibfnamefont {Z.}~\bibnamefont {Yuan}}, \bibinfo {author} {\bibfnamefont {M.-G.}\ \bibnamefont {Suh}},\ and\ \bibinfo {author} {\bibfnamefont {K.}~\bibnamefont {Vahala}},\ }\bibfield  {title} {\bibinfo {title} {{Petermann-factor sensitivity limit near an exceptional point in a Brillouin ring laser gyroscope}},\ }\href@noop {} {\bibfield  {journal} {\bibinfo  {journal} {Nat. Commun.}\ }\textbf {\bibinfo {volume} {11}},\ \bibinfo {pages} {1610} (\bibinfo {year} {2020})}\BibitemShut {NoStop}%
\bibitem [{\citenamefont {Ge}\ \emph {et~al.}(2010)\citenamefont {Ge}, \citenamefont {Chong},\ and\ \citenamefont {Stone}}]{Salt2010}%
  \BibitemOpen
  \bibfield  {author} {\bibinfo {author} {\bibfnamefont {L.}~\bibnamefont {Ge}}, \bibinfo {author} {\bibfnamefont {Y.~D.}\ \bibnamefont {Chong}},\ and\ \bibinfo {author} {\bibfnamefont {A.~D.}\ \bibnamefont {Stone}},\ }\bibfield  {title} {\bibinfo {title} {{Steady-state ab initio laser theory: Generalizations and analytic results}},\ }\href@noop {} {\bibfield  {journal} {\bibinfo  {journal} {Phys. Rev. A}\ }\textbf {\bibinfo {volume} {82}},\ \bibinfo {pages} {063824} (\bibinfo {year} {2010})}\BibitemShut {NoStop}%
\bibitem [{\citenamefont {Petermann}(1979)}]{petermann1979calculated}%
  \BibitemOpen
  \bibfield  {author} {\bibinfo {author} {\bibfnamefont {K.}~\bibnamefont {Petermann}},\ }\bibfield  {title} {\bibinfo {title} {{Calculated spontaneous emission factor for double-heterostructure injection lasers with gain-induced waveguiding}},\ }\href@noop {} {\bibfield  {journal} {\bibinfo  {journal} {IEEE J. Quantum Electron.}\ }\textbf {\bibinfo {volume} {15}},\ \bibinfo {pages} {566} (\bibinfo {year} {1979})}\BibitemShut {NoStop}%
\bibitem [{\citenamefont {Haus}\ and\ \citenamefont {Kawakami}(1985)}]{haus1985}%
  \BibitemOpen
  \bibfield  {author} {\bibinfo {author} {\bibfnamefont {H.}~\bibnamefont {Haus}}\ and\ \bibinfo {author} {\bibfnamefont {S.}~\bibnamefont {Kawakami}},\ }\bibfield  {title} {\bibinfo {title} {{On the $``$Excess spontaneous emission factor$''$ in gain-guided laser amplifiers}},\ }\href@noop {} {\bibfield  {journal} {\bibinfo  {journal} {IEEE J.~Quant.~Electr.}\ }\textbf {\bibinfo {volume} {21}},\ \bibinfo {pages} {63} (\bibinfo {year} {1985})}\BibitemShut {NoStop}%
\bibitem [{\citenamefont {Siegman}(1989{\natexlab{a}})}]{siegman1989excess1}%
  \BibitemOpen
  \bibfield  {author} {\bibinfo {author} {\bibfnamefont {A.~E.}\ \bibnamefont {Siegman}},\ }\bibfield  {title} {\bibinfo {title} {{Excess spontaneous emission in non-Hermitian optical systems. I. Laser amplifiers}},\ }\href@noop {} {\bibfield  {journal} {\bibinfo  {journal} {Phys. Rev. A}\ }\textbf {\bibinfo {volume} {39}},\ \bibinfo {pages} {1253} (\bibinfo {year} {1989}{\natexlab{a}})}\BibitemShut {NoStop}%
\bibitem [{\citenamefont {Siegman}(1989{\natexlab{b}})}]{siegman1989excess2}%
  \BibitemOpen
  \bibfield  {author} {\bibinfo {author} {\bibfnamefont {A.~E.}\ \bibnamefont {Siegman}},\ }\bibfield  {title} {\bibinfo {title} {{Excess spontaneous emission in non-Hermitian optical systems. II. Laser oscillators}},\ }\href@noop {} {\bibfield  {journal} {\bibinfo  {journal} {Phys. Rev. A}\ }\textbf {\bibinfo {volume} {39}},\ \bibinfo {pages} {1264} (\bibinfo {year} {1989}{\natexlab{b}})}\BibitemShut {NoStop}%
\bibitem [{\citenamefont {Wiersig}(2023)}]{wiersig2023petermann}%
  \BibitemOpen
  \bibfield  {author} {\bibinfo {author} {\bibfnamefont {J.}~\bibnamefont {Wiersig}},\ }\bibfield  {title} {\bibinfo {title} {{Petermann factors and phase rigidities near exceptional points}},\ }\href@noop {} {\bibfield  {journal} {\bibinfo  {journal} {Phys. Rev. Research}\ }\textbf {\bibinfo {volume} {5}},\ \bibinfo {pages} {033042} (\bibinfo {year} {2023})}\BibitemShut {NoStop}%
\bibitem [{\citenamefont {Peters}\ and\ \citenamefont {Rodriguez}(2022)}]{peters2022exceptional}%
  \BibitemOpen
  \bibfield  {author} {\bibinfo {author} {\bibfnamefont {K.~J.}\ \bibnamefont {Peters}}\ and\ \bibinfo {author} {\bibfnamefont {S.~R.~K.}\ \bibnamefont {Rodriguez}},\ }\bibfield  {title} {\bibinfo {title} {{Exceptional precision of a nonlinear optical sensor at a square-root singularity}},\ }\href@noop {} {\bibfield  {journal} {\bibinfo  {journal} {Phys. Rev. Lett.}\ }\textbf {\bibinfo {volume} {129}},\ \bibinfo {pages} {013901} (\bibinfo {year} {2022})}\BibitemShut {NoStop}%
\bibitem [{\citenamefont {Bai}\ \emph {et~al.}(2023{\natexlab{a}})\citenamefont {Bai}, \citenamefont {Li}, \citenamefont {Liu}, \citenamefont {Fang}, \citenamefont {Wan},\ and\ \citenamefont {Xiao}}]{bai2023nonlinear}%
  \BibitemOpen
  \bibfield  {author} {\bibinfo {author} {\bibfnamefont {K.}~\bibnamefont {Bai}}, \bibinfo {author} {\bibfnamefont {J.-Z.}\ \bibnamefont {Li}}, \bibinfo {author} {\bibfnamefont {T.-R.}\ \bibnamefont {Liu}}, \bibinfo {author} {\bibfnamefont {L.}~\bibnamefont {Fang}}, \bibinfo {author} {\bibfnamefont {D.}~\bibnamefont {Wan}},\ and\ \bibinfo {author} {\bibfnamefont {M.}~\bibnamefont {Xiao}},\ }\bibfield  {title} {\bibinfo {title} {{Nonlinear Exceptional Points with a Complete Basis in Dynamics}},\ }\href@noop {} {\bibfield  {journal} {\bibinfo  {journal} {Phys. Rev. Lett.}\ }\textbf {\bibinfo {volume} {130}},\ \bibinfo {pages} {266901} (\bibinfo {year} {2023}{\natexlab{a}})}\BibitemShut {NoStop}%
\bibitem [{\citenamefont {Bai}\ \emph {et~al.}(2023{\natexlab{b}})\citenamefont {Bai}, \citenamefont {Fang}, \citenamefont {Liu}, \citenamefont {Li}, \citenamefont {Wan},\ and\ \citenamefont {Xiao}}]{bai2023nonlinearity}%
  \BibitemOpen
  \bibfield  {author} {\bibinfo {author} {\bibfnamefont {K.}~\bibnamefont {Bai}}, \bibinfo {author} {\bibfnamefont {L.}~\bibnamefont {Fang}}, \bibinfo {author} {\bibfnamefont {T.-R.}\ \bibnamefont {Liu}}, \bibinfo {author} {\bibfnamefont {J.-Z.}\ \bibnamefont {Li}}, \bibinfo {author} {\bibfnamefont {D.}~\bibnamefont {Wan}},\ and\ \bibinfo {author} {\bibfnamefont {M.}~\bibnamefont {Xiao}},\ }\bibfield  {title} {\bibinfo {title} {{Nonlinearity-enabled higher-order exceptional singularities with ultra-enhanced signal-to-noise ratio}},\ }\href@noop {} {\bibfield  {journal} {\bibinfo  {journal} {Nat. Sci. Rev.}\ }\textbf {\bibinfo {volume} {10}},\ \bibinfo {pages} {nwac259} (\bibinfo {year} {2023}{\natexlab{b}})}\BibitemShut {NoStop}%
\bibitem [{\citenamefont {Bai}\ \emph {et~al.}(2024)\citenamefont {Bai}, \citenamefont {Liu}, \citenamefont {Fang}, \citenamefont {Li}, \citenamefont {Lin}, \citenamefont {Wan},\ and\ \citenamefont {Xiao}}]{bai2024observation}%
  \BibitemOpen
  \bibfield  {author} {\bibinfo {author} {\bibfnamefont {K.}~\bibnamefont {Bai}}, \bibinfo {author} {\bibfnamefont {T.-R.}\ \bibnamefont {Liu}}, \bibinfo {author} {\bibfnamefont {L.}~\bibnamefont {Fang}}, \bibinfo {author} {\bibfnamefont {J.-Z.}\ \bibnamefont {Li}}, \bibinfo {author} {\bibfnamefont {C.}~\bibnamefont {Lin}}, \bibinfo {author} {\bibfnamefont {D.}~\bibnamefont {Wan}},\ and\ \bibinfo {author} {\bibfnamefont {M.}~\bibnamefont {Xiao}},\ }\bibfield  {title} {\bibinfo {title} {Observation of nonlinear exceptional points with a complete basis in dynamics},\ }\href@noop {} {\bibfield  {journal} {\bibinfo  {journal} {Phys. Rev. Lett.}\ }\textbf {\bibinfo {volume} {132}},\ \bibinfo {pages} {073802} (\bibinfo {year} {2024})}\BibitemShut {NoStop}%
\bibitem [{\citenamefont {Holmes}\ and\ \citenamefont {Rand}(1978)}]{holmes1978bifurcations}%
  \BibitemOpen
  \bibfield  {author} {\bibinfo {author} {\bibfnamefont {P.}~\bibnamefont {Holmes}}\ and\ \bibinfo {author} {\bibfnamefont {D.}~\bibnamefont {Rand}},\ }\bibfield  {title} {\bibinfo {title} {{Bifurcations of the Forced Van der Pol Oscillator}},\ }\href@noop {} {\bibfield  {journal} {\bibinfo  {journal} {Q. Appl. Math.}\ }\textbf {\bibinfo {volume} {35}},\ \bibinfo {pages} {495} (\bibinfo {year} {1978})}\BibitemShut {NoStop}%
\bibitem [{\citenamefont {Lee}\ and\ \citenamefont {Sadeghpour}(2013)}]{lee2013quantum}%
  \BibitemOpen
  \bibfield  {author} {\bibinfo {author} {\bibfnamefont {T.~E.}\ \bibnamefont {Lee}}\ and\ \bibinfo {author} {\bibfnamefont {H.~R.}\ \bibnamefont {Sadeghpour}},\ }\bibfield  {title} {\bibinfo {title} {{Quantum Synchronization of Quantum Van der Pol Oscillators with Trapped Ions}},\ }\href@noop {} {\bibfield  {journal} {\bibinfo  {journal} {Phys. Rev. Lett.}\ }\textbf {\bibinfo {volume} {111}},\ \bibinfo {pages} {234101} (\bibinfo {year} {2013})}\BibitemShut {NoStop}%
\bibitem [{\citenamefont {Walter}\ \emph {et~al.}(2015)\citenamefont {Walter}, \citenamefont {Nunnenkamp},\ and\ \citenamefont {Bruder}}]{walter2015quantum}%
  \BibitemOpen
  \bibfield  {author} {\bibinfo {author} {\bibfnamefont {S.}~\bibnamefont {Walter}}, \bibinfo {author} {\bibfnamefont {A.}~\bibnamefont {Nunnenkamp}},\ and\ \bibinfo {author} {\bibfnamefont {C.}~\bibnamefont {Bruder}},\ }\bibfield  {title} {\bibinfo {title} {{Quantum Synchronization of Two Van der Pol Oscillators}},\ }\href@noop {} {\bibfield  {journal} {\bibinfo  {journal} {Ann. Phys.}\ }\textbf {\bibinfo {volume} {527}},\ \bibinfo {pages} {131} (\bibinfo {year} {2015})}\BibitemShut {NoStop}%
\bibitem [{\citenamefont {Dutta}\ and\ \citenamefont {Cooper}(2019)}]{dutta2019critical}%
  \BibitemOpen
  \bibfield  {author} {\bibinfo {author} {\bibfnamefont {S.}~\bibnamefont {Dutta}}\ and\ \bibinfo {author} {\bibfnamefont {N.~R.}\ \bibnamefont {Cooper}},\ }\bibfield  {title} {\bibinfo {title} {{Critical Response of a Quantum Van der Pol Oscillator}},\ }\href@noop {} {\bibfield  {journal} {\bibinfo  {journal} {Phys. Rev. Lett.}\ }\textbf {\bibinfo {volume} {123}},\ \bibinfo {pages} {250401} (\bibinfo {year} {2019})}\BibitemShut {NoStop}%
\bibitem [{sup()}]{supplement}%
  \BibitemOpen
  \href@noop {} {\bibinfo {title} {See supplemental material, which includes {R}efs. \cite{gardiner2004quantum,scully1997quantum,cramer1999mathematical,braunstein1994statistical,paris2009quantum} , for additional information about detailed discussions of the analytic methods, numerical simulations and potential experimental platforms.}}\BibitemShut {Stop}%
\bibitem [{\citenamefont {Chow}\ \emph {et~al.}(1985)\citenamefont {Chow}, \citenamefont {Gea-Banacloche}, \citenamefont {Pedrotti}, \citenamefont {Sanders}, \citenamefont {Schleich},\ and\ \citenamefont {Scully}}]{chow1985ring}%
  \BibitemOpen
  \bibfield  {author} {\bibinfo {author} {\bibfnamefont {W.}~\bibnamefont {Chow}}, \bibinfo {author} {\bibfnamefont {J.}~\bibnamefont {Gea-Banacloche}}, \bibinfo {author} {\bibfnamefont {L.}~\bibnamefont {Pedrotti}}, \bibinfo {author} {\bibfnamefont {V.}~\bibnamefont {Sanders}}, \bibinfo {author} {\bibfnamefont {W.}~\bibnamefont {Schleich}},\ and\ \bibinfo {author} {\bibfnamefont {M.}~\bibnamefont {Scully}},\ }\bibfield  {title} {\bibinfo {title} {The ring laser gyro},\ }\href@noop {} {\bibfield  {journal} {\bibinfo  {journal} {Rev. Mod. Phys.}\ }\textbf {\bibinfo {volume} {57}},\ \bibinfo {pages} {61} (\bibinfo {year} {1985})}\BibitemShut {NoStop}%
\bibitem [{\citenamefont {Lee}\ \emph {et~al.}(2008)\citenamefont {Lee}, \citenamefont {Ryu}, \citenamefont {Shim}, \citenamefont {Lee}, \citenamefont {Kim},\ and\ \citenamefont {An}}]{lee2008divergent}%
  \BibitemOpen
  \bibfield  {author} {\bibinfo {author} {\bibfnamefont {S.-Y.}\ \bibnamefont {Lee}}, \bibinfo {author} {\bibfnamefont {J.-W.}\ \bibnamefont {Ryu}}, \bibinfo {author} {\bibfnamefont {J.-B.}\ \bibnamefont {Shim}}, \bibinfo {author} {\bibfnamefont {S.-B.}\ \bibnamefont {Lee}}, \bibinfo {author} {\bibfnamefont {S.~W.}\ \bibnamefont {Kim}},\ and\ \bibinfo {author} {\bibfnamefont {K.}~\bibnamefont {An}},\ }\bibfield  {title} {\bibinfo {title} {{Divergent Petermann factor of interacting resonances in a stadium-shaped microcavity}},\ }\href@noop {} {\bibfield  {journal} {\bibinfo  {journal} {Phys. Rev. A}\ }\textbf {\bibinfo {volume} {78}},\ \bibinfo {pages} {015805} (\bibinfo {year} {2008})}\BibitemShut {NoStop}%
\bibitem [{\citenamefont {Smith}\ \emph {et~al.}(2022)\citenamefont {Smith}, \citenamefont {Chang}, \citenamefont {Mikhailov},\ and\ \citenamefont {Shahriar}}]{smith2022beyond}%
  \BibitemOpen
  \bibfield  {author} {\bibinfo {author} {\bibfnamefont {D.~D.}\ \bibnamefont {Smith}}, \bibinfo {author} {\bibfnamefont {H.}~\bibnamefont {Chang}}, \bibinfo {author} {\bibfnamefont {E.}~\bibnamefont {Mikhailov}},\ and\ \bibinfo {author} {\bibfnamefont {S.~M.}\ \bibnamefont {Shahriar}},\ }\bibfield  {title} {\bibinfo {title} {{Beyond the Petermann limit: Prospect of increasing sensor precision near exceptional points}},\ }\href@noop {} {\bibfield  {journal} {\bibinfo  {journal} {Phys. Rev. A}\ }\textbf {\bibinfo {volume} {106}},\ \bibinfo {pages} {013520} (\bibinfo {year} {2022})}\BibitemShut {NoStop}%
\bibitem [{\citenamefont {Rackauckas}\ and\ \citenamefont {Nie}(2017)}]{rackauckas2017differentialequations}%
  \BibitemOpen
  \bibfield  {author} {\bibinfo {author} {\bibfnamefont {C.}~\bibnamefont {Rackauckas}}\ and\ \bibinfo {author} {\bibfnamefont {Q.}~\bibnamefont {Nie}},\ }\bibfield  {title} {\bibinfo {title} {Differential{E}quations.jl--a performant and feature-rich ecosystem for solving differential equations in {J}ulia},\ }\href@noop {} {\bibfield  {journal} {\bibinfo  {journal} {J. Open Res. Softw.}\ }\textbf {\bibinfo {volume} {5}} (\bibinfo {year} {2017})}\BibitemShut {NoStop}%
\bibitem [{\citenamefont {R{\"o}{\ss}ler}(2009)}]{rossler2009second}%
  \BibitemOpen
  \bibfield  {author} {\bibinfo {author} {\bibfnamefont {A.}~\bibnamefont {R{\"o}{\ss}ler}},\ }\bibfield  {title} {\bibinfo {title} {{Second order Runge--Kutta methods for It{\^o} stochastic differential equations}},\ }\href@noop {} {\bibfield  {journal} {\bibinfo  {journal} {SIAM J. Numer. Anal.}\ }\textbf {\bibinfo {volume} {47}},\ \bibinfo {pages} {1713} (\bibinfo {year} {2009})}\BibitemShut {NoStop}%
\bibitem [{\citenamefont {Van~den Broeck}\ \emph {et~al.}(1994)\citenamefont {Van~den Broeck}, \citenamefont {Parrondo},\ and\ \citenamefont {Toral}}]{van1994noise}%
  \BibitemOpen
  \bibfield  {author} {\bibinfo {author} {\bibfnamefont {C.}~\bibnamefont {Van~den Broeck}}, \bibinfo {author} {\bibfnamefont {J.~M.~R.}\ \bibnamefont {Parrondo}},\ and\ \bibinfo {author} {\bibfnamefont {R.}~\bibnamefont {Toral}},\ }\bibfield  {title} {\bibinfo {title} {Noise-induced nonequilibrium phase transition},\ }\href@noop {} {\bibfield  {journal} {\bibinfo  {journal} {Phys. Rev. Lett.}\ }\textbf {\bibinfo {volume} {73}},\ \bibinfo {pages} {3395} (\bibinfo {year} {1994})}\BibitemShut {NoStop}%
\bibitem [{\citenamefont {{St-Jean}}\ \emph {et~al.}(2017)\citenamefont {{St-Jean}}, \citenamefont {Goblot}, \citenamefont {Galopin}, \citenamefont {Lema{\^i}tre}, \citenamefont {Ozawa}, \citenamefont {Le~Gratiet}, \citenamefont {Sagnes}, \citenamefont {Bloch},\ and\ \citenamefont {Amo}}]{st-jeanLasingTopologicalEdge2017}%
  \BibitemOpen
  \bibfield  {author} {\bibinfo {author} {\bibfnamefont {P.}~\bibnamefont {{St-Jean}}}, \bibinfo {author} {\bibfnamefont {V.}~\bibnamefont {Goblot}}, \bibinfo {author} {\bibfnamefont {E.}~\bibnamefont {Galopin}}, \bibinfo {author} {\bibfnamefont {A.}~\bibnamefont {Lema{\^i}tre}}, \bibinfo {author} {\bibfnamefont {T.}~\bibnamefont {Ozawa}}, \bibinfo {author} {\bibfnamefont {L.}~\bibnamefont {Le~Gratiet}}, \bibinfo {author} {\bibfnamefont {I.}~\bibnamefont {Sagnes}}, \bibinfo {author} {\bibfnamefont {J.}~\bibnamefont {Bloch}},\ and\ \bibinfo {author} {\bibfnamefont {A.}~\bibnamefont {Amo}},\ }\bibfield  {title} {\bibinfo {title} {Lasing in topological edge states of a one-dimensional lattice},\ }\href@noop {} {\bibfield  {journal} {\bibinfo  {journal} {Nat. Photonics}\ }\textbf {\bibinfo {volume} {11}},\ \bibinfo {pages} {651} (\bibinfo {year} {2017})}\BibitemShut {NoStop}%
\bibitem [{\citenamefont {Parto}\ \emph {et~al.}(2018)\citenamefont {Parto}, \citenamefont {Wittek}, \citenamefont {Hodaei}, \citenamefont {Harari}, \citenamefont {Bandres}, \citenamefont {Ren}, \citenamefont {Rechtsman}, \citenamefont {Segev}, \citenamefont {Christodoulides},\ and\ \citenamefont {Khajavikhan}}]{partoEdgeModeLasing1D2018}%
  \BibitemOpen
  \bibfield  {author} {\bibinfo {author} {\bibfnamefont {M.}~\bibnamefont {Parto}}, \bibinfo {author} {\bibfnamefont {S.}~\bibnamefont {Wittek}}, \bibinfo {author} {\bibfnamefont {H.}~\bibnamefont {Hodaei}}, \bibinfo {author} {\bibfnamefont {G.}~\bibnamefont {Harari}}, \bibinfo {author} {\bibfnamefont {M.~A.}\ \bibnamefont {Bandres}}, \bibinfo {author} {\bibfnamefont {J.}~\bibnamefont {Ren}}, \bibinfo {author} {\bibfnamefont {M.~C.}\ \bibnamefont {Rechtsman}}, \bibinfo {author} {\bibfnamefont {M.}~\bibnamefont {Segev}}, \bibinfo {author} {\bibfnamefont {D.~N.}\ \bibnamefont {Christodoulides}},\ and\ \bibinfo {author} {\bibfnamefont {M.}~\bibnamefont {Khajavikhan}},\ }\bibfield  {title} {\bibinfo {title} {Edge-{{Mode Lasing}} in {{1D Topological Active Arrays}}},\ }\href@noop {} {\bibfield  {journal} {\bibinfo  {journal} {Phys. Rev. Lett.}\ }\textbf {\bibinfo {volume} {120}},\ \bibinfo {pages} {113901} (\bibinfo {year} {2018})}\BibitemShut {NoStop}%
\bibitem [{\citenamefont {Zhao}\ \emph {et~al.}(2018)\citenamefont {Zhao}, \citenamefont {Miao}, \citenamefont {Teimourpour}, \citenamefont {Malzard}, \citenamefont {{El-Ganainy}}, \citenamefont {Schomerus},\ and\ \citenamefont {Feng}}]{zhaoTopologicalHybridSilicon2018}%
  \BibitemOpen
  \bibfield  {author} {\bibinfo {author} {\bibfnamefont {H.}~\bibnamefont {Zhao}}, \bibinfo {author} {\bibfnamefont {P.}~\bibnamefont {Miao}}, \bibinfo {author} {\bibfnamefont {M.~H.}\ \bibnamefont {Teimourpour}}, \bibinfo {author} {\bibfnamefont {S.}~\bibnamefont {Malzard}}, \bibinfo {author} {\bibfnamefont {R.}~\bibnamefont {{El-Ganainy}}}, \bibinfo {author} {\bibfnamefont {H.}~\bibnamefont {Schomerus}},\ and\ \bibinfo {author} {\bibfnamefont {L.}~\bibnamefont {Feng}},\ }\bibfield  {title} {\bibinfo {title} {Topological hybrid silicon microlasers},\ }\href@noop {} {\bibfield  {journal} {\bibinfo  {journal} {Nat. Commun.}\ }\textbf {\bibinfo {volume} {9}},\ \bibinfo {pages} {981} (\bibinfo {year} {2018})}\BibitemShut {NoStop}%
\bibitem [{\citenamefont {Bandres}\ \emph {et~al.}(2018)\citenamefont {Bandres}, \citenamefont {Wittek}, \citenamefont {Harari}, \citenamefont {Parto}, \citenamefont {Ren}, \citenamefont {Segev}, \citenamefont {Christodoulides},\ and\ \citenamefont {Khajavikhan}}]{bandresTopologicalInsulatorLaser2018}%
  \BibitemOpen
  \bibfield  {author} {\bibinfo {author} {\bibfnamefont {M.~A.}\ \bibnamefont {Bandres}}, \bibinfo {author} {\bibfnamefont {S.}~\bibnamefont {Wittek}}, \bibinfo {author} {\bibfnamefont {G.}~\bibnamefont {Harari}}, \bibinfo {author} {\bibfnamefont {M.}~\bibnamefont {Parto}}, \bibinfo {author} {\bibfnamefont {J.}~\bibnamefont {Ren}}, \bibinfo {author} {\bibfnamefont {M.}~\bibnamefont {Segev}}, \bibinfo {author} {\bibfnamefont {D.~N.}\ \bibnamefont {Christodoulides}},\ and\ \bibinfo {author} {\bibfnamefont {M.}~\bibnamefont {Khajavikhan}},\ }\bibfield  {title} {\bibinfo {title} {Topological insulator laser: {{Experiments}}},\ }\href@noop {} {\bibfield  {journal} {\bibinfo  {journal} {Science}\ }\textbf {\bibinfo {volume} {359}},\ \bibinfo {pages} {eaar4005} (\bibinfo {year} {2018})}\BibitemShut {NoStop}%
\bibitem [{\citenamefont {Peng}\ \emph {et~al.}(2014)\citenamefont {Peng}, \citenamefont {{\"O}zdemir}, \citenamefont {Rotter}, \citenamefont {Yilmaz}, \citenamefont {Liertzer}, \citenamefont {Monifi}, \citenamefont {Bender}, \citenamefont {Nori},\ and\ \citenamefont {Yang}}]{pengLossinducedSuppressionRevival2014}%
  \BibitemOpen
  \bibfield  {author} {\bibinfo {author} {\bibfnamefont {B.}~\bibnamefont {Peng}}, \bibinfo {author} {\bibfnamefont {{\c S}.~K.}\ \bibnamefont {{\"O}zdemir}}, \bibinfo {author} {\bibfnamefont {S.}~\bibnamefont {Rotter}}, \bibinfo {author} {\bibfnamefont {H.}~\bibnamefont {Yilmaz}}, \bibinfo {author} {\bibfnamefont {M.}~\bibnamefont {Liertzer}}, \bibinfo {author} {\bibfnamefont {F.}~\bibnamefont {Monifi}}, \bibinfo {author} {\bibfnamefont {C.~M.}\ \bibnamefont {Bender}}, \bibinfo {author} {\bibfnamefont {F.}~\bibnamefont {Nori}},\ and\ \bibinfo {author} {\bibfnamefont {L.}~\bibnamefont {Yang}},\ }\bibfield  {title} {\bibinfo {title} {Loss-induced suppression and revival of lasing},\ }\href@noop {} {\bibfield  {journal} {\bibinfo  {journal} {Science}\ }\textbf {\bibinfo {volume} {346}},\ \bibinfo {pages} {328} (\bibinfo {year} {2014})}\BibitemShut {NoStop}%
\bibitem [{\citenamefont {Ramezanpour}\ and\ \citenamefont {Bogdanov}(2021)}]{ramezanpour2021tuning}%
  \BibitemOpen
  \bibfield  {author} {\bibinfo {author} {\bibfnamefont {S.}~\bibnamefont {Ramezanpour}}\ and\ \bibinfo {author} {\bibfnamefont {A.}~\bibnamefont {Bogdanov}},\ }\bibfield  {title} {\bibinfo {title} {{Tuning Exceptional Points with Kerr Nonlinearity}},\ }\href@noop {} {\bibfield  {journal} {\bibinfo  {journal} {Phys. Rev. A}\ }\textbf {\bibinfo {volume} {103}},\ \bibinfo {pages} {043510} (\bibinfo {year} {2021})}\BibitemShut {NoStop}%
\bibitem [{\citenamefont {Kononchuk}\ \emph {et~al.}(2022)\citenamefont {Kononchuk}, \citenamefont {Cai}, \citenamefont {Ellis}, \citenamefont {Thevamaran},\ and\ \citenamefont {Kottos}}]{kononchuk2022exceptional}%
  \BibitemOpen
  \bibfield  {author} {\bibinfo {author} {\bibfnamefont {R.}~\bibnamefont {Kononchuk}}, \bibinfo {author} {\bibfnamefont {J.}~\bibnamefont {Cai}}, \bibinfo {author} {\bibfnamefont {F.}~\bibnamefont {Ellis}}, \bibinfo {author} {\bibfnamefont {R.}~\bibnamefont {Thevamaran}},\ and\ \bibinfo {author} {\bibfnamefont {T.}~\bibnamefont {Kottos}},\ }\bibfield  {title} {\bibinfo {title} {{Exceptional-Point-Based Accelerometers with Enhanced Signal-to-Noise Ratio}},\ }\href@noop {} {\bibfield  {journal} {\bibinfo  {journal} {Nature}\ }\textbf {\bibinfo {volume} {607}},\ \bibinfo {pages} {697} (\bibinfo {year} {2022})}\BibitemShut {NoStop}%
\bibitem [{\citenamefont {Degen}\ \emph {et~al.}(2017)\citenamefont {Degen}, \citenamefont {Reinhard},\ and\ \citenamefont {Cappellaro}}]{degen2017quantum}%
  \BibitemOpen
  \bibfield  {author} {\bibinfo {author} {\bibfnamefont {C.~L.}\ \bibnamefont {Degen}}, \bibinfo {author} {\bibfnamefont {F.}~\bibnamefont {Reinhard}},\ and\ \bibinfo {author} {\bibfnamefont {P.}~\bibnamefont {Cappellaro}},\ }\bibfield  {title} {\bibinfo {title} {{Quantum sensing}},\ }\href@noop {} {\bibfield  {journal} {\bibinfo  {journal} {Rev. Mod. Phys.}\ }\textbf {\bibinfo {volume} {89}},\ \bibinfo {pages} {035002} (\bibinfo {year} {2017})}\BibitemShut {NoStop}%
\bibitem [{\citenamefont {Pirandola}\ \emph {et~al.}(2018)\citenamefont {Pirandola}, \citenamefont {Bardhan}, \citenamefont {Gehring}, \citenamefont {Weedbrook},\ and\ \citenamefont {Lloyd}}]{pirandola2018advances}%
  \BibitemOpen
  \bibfield  {author} {\bibinfo {author} {\bibfnamefont {S.}~\bibnamefont {Pirandola}}, \bibinfo {author} {\bibfnamefont {B.~R.}\ \bibnamefont {Bardhan}}, \bibinfo {author} {\bibfnamefont {T.}~\bibnamefont {Gehring}}, \bibinfo {author} {\bibfnamefont {C.}~\bibnamefont {Weedbrook}},\ and\ \bibinfo {author} {\bibfnamefont {S.}~\bibnamefont {Lloyd}},\ }\bibfield  {title} {\bibinfo {title} {{Advances in photonic quantum sensing}},\ }\href@noop {} {\bibfield  {journal} {\bibinfo  {journal} {Nat. Photonics}\ }\textbf {\bibinfo {volume} {12}},\ \bibinfo {pages} {724} (\bibinfo {year} {2018})}\BibitemShut {NoStop}%
\bibitem [{\citenamefont {Lawrie}\ \emph {et~al.}(2019)\citenamefont {Lawrie}, \citenamefont {Lett}, \citenamefont {Marino},\ and\ \citenamefont {Pooser}}]{lawrie2019quantum}%
  \BibitemOpen
  \bibfield  {author} {\bibinfo {author} {\bibfnamefont {B.~J.}\ \bibnamefont {Lawrie}}, \bibinfo {author} {\bibfnamefont {P.~D.}\ \bibnamefont {Lett}}, \bibinfo {author} {\bibfnamefont {A.~M.}\ \bibnamefont {Marino}},\ and\ \bibinfo {author} {\bibfnamefont {R.~C.}\ \bibnamefont {Pooser}},\ }\bibfield  {title} {\bibinfo {title} {{Quantum sensing with squeezed light}},\ }\href@noop {} {\bibfield  {journal} {\bibinfo  {journal} {ACS Photonics}\ }\textbf {\bibinfo {volume} {6}},\ \bibinfo {pages} {1307} (\bibinfo {year} {2019})}\BibitemShut {NoStop}%
\bibitem [{\citenamefont {Aslam}\ \emph {et~al.}(2023)\citenamefont {Aslam}, \citenamefont {Zhou}, \citenamefont {Urbach}, \citenamefont {Turner}, \citenamefont {Walsworth}, \citenamefont {Lukin},\ and\ \citenamefont {Park}}]{aslam2023quantum}%
  \BibitemOpen
  \bibfield  {author} {\bibinfo {author} {\bibfnamefont {N.}~\bibnamefont {Aslam}}, \bibinfo {author} {\bibfnamefont {H.}~\bibnamefont {Zhou}}, \bibinfo {author} {\bibfnamefont {E.~K.}\ \bibnamefont {Urbach}}, \bibinfo {author} {\bibfnamefont {M.~J.}\ \bibnamefont {Turner}}, \bibinfo {author} {\bibfnamefont {R.~L.}\ \bibnamefont {Walsworth}}, \bibinfo {author} {\bibfnamefont {M.~D.}\ \bibnamefont {Lukin}},\ and\ \bibinfo {author} {\bibfnamefont {H.}~\bibnamefont {Park}},\ }\bibfield  {title} {\bibinfo {title} {{Quantum sensors for biomedical applications}},\ }\href@noop {} {\bibfield  {journal} {\bibinfo  {journal} {Nat. Rev. Phys.}\ }\textbf {\bibinfo {volume} {5}},\ \bibinfo {pages} {157} (\bibinfo {year} {2023})}\BibitemShut {NoStop}%
\bibitem [{\citenamefont {Bass}\ and\ \citenamefont {Doser}(2024)}]{bass2024quantum}%
  \BibitemOpen
  \bibfield  {author} {\bibinfo {author} {\bibfnamefont {S.~D.}\ \bibnamefont {Bass}}\ and\ \bibinfo {author} {\bibfnamefont {M.}~\bibnamefont {Doser}},\ }\bibfield  {title} {\bibinfo {title} {{Quantum sensing for particle physics}},\ }\href@noop {} {\bibfield  {journal} {\bibinfo  {journal} {Nat. Rev. Phys.}\ }\textbf {\bibinfo {volume} {6}},\ \bibinfo {pages} {329–339} (\bibinfo {year} {2024})}\BibitemShut {NoStop}%
\bibitem [{\citenamefont {Gardiner}\ and\ \citenamefont {Zoller}(2004)}]{gardiner2004quantum}%
  \BibitemOpen
  \bibfield  {author} {\bibinfo {author} {\bibfnamefont {C.}~\bibnamefont {Gardiner}}\ and\ \bibinfo {author} {\bibfnamefont {P.}~\bibnamefont {Zoller}},\ }\href@noop {} {\emph {\bibinfo {title} {{Quantum noise: a handbook of Markovian and non-Markovian quantum stochastic methods with applications to quantum optics}}}}\ (\bibinfo  {publisher} {Springer Science \& Business Media},\ \bibinfo {year} {2004})\BibitemShut {NoStop}%
\bibitem [{\citenamefont {Scully}\ and\ \citenamefont {Zubairy}(1997)}]{scully1997quantum}%
  \BibitemOpen
  \bibfield  {author} {\bibinfo {author} {\bibfnamefont {M.~O.}\ \bibnamefont {Scully}}\ and\ \bibinfo {author} {\bibfnamefont {M.~S.}\ \bibnamefont {Zubairy}},\ }\href@noop {} {\emph {\bibinfo {title} {{Quantum optics}}}}\ (\bibinfo  {publisher} {Cambridge University Press},\ \bibinfo {year} {1997})\BibitemShut {NoStop}%
\bibitem [{\citenamefont {Cram{\'e}r}(1999)}]{cramer1999mathematical}%
  \BibitemOpen
  \bibfield  {author} {\bibinfo {author} {\bibfnamefont {H.}~\bibnamefont {Cram{\'e}r}},\ }\href@noop {} {\emph {\bibinfo {title} {{Mathematical methods of statistics}}}},\ Vol.~\bibinfo {volume} {26}\ (\bibinfo  {publisher} {Princeton university press},\ \bibinfo {year} {1999})\BibitemShut {NoStop}%
\bibitem [{\citenamefont {Braunstein}\ and\ \citenamefont {Caves}(1994)}]{braunstein1994statistical}%
  \BibitemOpen
  \bibfield  {author} {\bibinfo {author} {\bibfnamefont {S.~L.}\ \bibnamefont {Braunstein}}\ and\ \bibinfo {author} {\bibfnamefont {C.~M.}\ \bibnamefont {Caves}},\ }\bibfield  {title} {\bibinfo {title} {{Statistical distance and the geometry of quantum states}},\ }\href@noop {} {\bibfield  {journal} {\bibinfo  {journal} {Phys. Rev. Lett.}\ }\textbf {\bibinfo {volume} {72}},\ \bibinfo {pages} {3439} (\bibinfo {year} {1994})}\BibitemShut {NoStop}%
\bibitem [{\citenamefont {Paris}(2009)}]{paris2009quantum}%
  \BibitemOpen
  \bibfield  {author} {\bibinfo {author} {\bibfnamefont {M.~G.}\ \bibnamefont {Paris}},\ }\bibfield  {title} {\bibinfo {title} {{Quantum estimation for quantum technology}},\ }\href@noop {} {\bibfield  {journal} {\bibinfo  {journal} {Int. J. Quant. Inf.}\ }\textbf {\bibinfo {volume} {7}},\ \bibinfo {pages} {125} (\bibinfo {year} {2009})}\BibitemShut {NoStop}%
\end{thebibliography}%

\begin{widetext}
\newpage
  
\makeatletter 
\renewcommand{\theequation}{S\arabic{equation}}
\makeatother
\setcounter{equation}{0}

\renewcommand{\thesection}{S\arabic{section}} 
\setcounter{section}{0}

\makeatletter 
\renewcommand{\thefigure}{S\@arabic\c@figure}
\makeatother
\setcounter{figure}{0}

\begin{center}
  \textbf{Supplemental Materials for}\\
  \vskip 0.1in
  {\large ``Noise Constraints for Nonlinear Exceptional Point Sensing''}\\
  \vskip 0.1in
  {\small Xu Zheng and Y.~D.~Chong}
\end{center}
  
\section{Derivation of c-number Langevin equations for Van der Pol model}

In this section, we discuss how the $c$-number Langevin equations~(1)--(2) of the main text can emerge from a quantum van der Pol oscillator model.  We start from the quantum master equation \cite{holmes1978bifurcations, lee2013quantum, walter2015quantum, dutta2019critical}
\begin{align}
\dot{\rho} &= -i[\hat{H},\rho] + \gamma_a\mathcal{L}[\hat{a}^\dagger]\rho + \Gamma\mathcal{L}[\hat{a}^2]\rho + \gamma_b\mathcal{L}[\hat{b}]\rho,
\label{eq:mastereq}
\end{align}
where $\rho$ is the density matrix, $\hat{a}, \hat{b}$ are lowering operators for the $a$ and $b$ resonators respectively, $\gamma_a$ is the linear gain of the $a$ cavity, $\gamma_b$ is the loss rate of the $b$ cavity, and $\Gamma$ is the nonlinear loss of the $a$ cavity.  The Hamiltonian and Lindblad operators are given by
\begin{align}
&\hat{H} = \omega_a\hat{a}^\dagger\hat{a} + \omega_b\hat{b}^\dagger\hat{b} + g(\hat{a}^\dagger\hat{b} + \hat{a}\hat{b}^\dagger), \\
&\mathcal{L}[\hat{o}]\rho = \hat{o}\rho\hat{o}^\dagger - (\hat{o}^\dagger\hat{o}\rho + \rho\hat{o}^\dagger\hat{o})/2,
\end{align}
where $\omega_{a}, \omega_{b}$ are the natural frequencies of the two resonators and $g$ is the inter-resonator coupling. The Lindblad operator for $\hat{a}^2$ represents a nonlinear loss process.  Such a combination of linear gain and nonlinear loss effectively describes saturable gain.  To derive the corresponding Heisenberg-Langevin equations, we exploit the equivalence between the master equation and the Heisenberg-Langevin formalism \cite{gardiner2004quantum}:
\begin{align}
\dot{\hat{a}} &= \left(-i\omega_a + \frac{\gamma_a}{2}\right)\hat{a} - \Gamma\hat{a}^\dagger\hat{a}\hat{a} - ig\hat{b} + \hat{\xi}_a, \\
\dot{\hat{b}} &= \left(-i\omega_b - \frac{\gamma_b}{2}\right)\hat{b} - ig\hat{a} + \hat{\xi}_b.
\label{eq:quantumlangevineq}
\end{align}
The quantum noise operators $\hat{\xi}_{a,b}$ have zero mean and second-order correlations given by
\begin{align}
\langle\hat{\xi}_{\mu}(t)\hat{\xi}_{\nu}(t^{\prime})\rangle = \tilde{D}_{\mu\nu}\delta(t-t^{\prime}),
\label{eq:quantumnoise}
\end{align}
where the diffusion elements $\tilde{D}_{\mu\nu}$ are given by
\begin{align}
    \tilde{D}_{a^{\dagger}a} = \gamma_a,~ \tilde{D}_{aa^{\dagger}} = 4\Gamma\langle\hat{a}^\dagger\hat{a}\rangle,~ \tilde{D}_{bb^{\dagger}} = \gamma_b,
\end{align}
and all other elements are zero. 

To convert the operator equations to $c$-number equations, we must define an ordering for the operators (this is necessary since the $c$-numbers commute with each other, while the operators do not).  We choose the symmetric ordering and establish the mapping
\begin{align}
    (\hat{S}_j\hat{S}_k)_{\text{sym}}&=\frac{1}{2}\left(\hat{S}_j\hat{S}_k+\hat{S}_k\hat{S}_j\right)\rightarrow S_jS_k, \nonumber\\
    (\hat{S}_i\hat{S}_j\hat{S}_k)_{\text{sym}}&=\frac{1}{3!}\sum_{\sigma\in G}\hat{S}_{\sigma(1)}\hat{S}_{\sigma(2)}\hat{S}_{\sigma(3)}\rightarrow S_iS_jS_k,
\end{align}
where $G$ denotes the permutation group of the set $M=\{\hat{S}_i,\hat{S}_j,\hat{S}_k\}$, $\sigma$ is a permutation of the set $M$. Under symmetric ordering, the diffusion matrix satisfies the mapping
\begin{align}
    (\tilde{D}_{\mu\nu})_{\text{sym}}=\frac{1}{2}(\tilde{D}_{\mu\nu}+\tilde{D}_{\nu\mu})\rightarrow D_{\mu\nu}^c.
\end{align}
Following these mapping rules, we obtain the $c$-number Langevin equations
\begin{align}
    \dot{\alpha} &= \left[-i\omega_a + \frac{\gamma_a}{2}+\Gamma(1-|\alpha|^2)\right]\alpha - ig\beta + \xi_\alpha,
    \label{eq:classicallangevineq1} \\
    \dot{\beta} &= \left(-i\omega_b - \frac{\gamma_b}{2}\right)\beta - ig\alpha + \xi_\beta.
    \label{eq:classicallangevineq2}
\end{align}
The noise terms $\xi_{\alpha,\beta}$ have zero mean and have second-order correlations satisfying
\begin{align}
\langle\xi_{\mu}(t)\xi_{\nu}(t')\rangle=D_{\mu\nu}^c\delta(t-t'),
\label{eq:classicalnoise}
\end{align}
where
\begin{align}
  D_{\alpha^{\ast}\alpha}^c &= D_{\alpha\alpha^{\ast}}^c
  = D\left[\frac{\gamma_a}{2} + \Gamma\left(2\langle|\alpha|^2\rangle - 1\right)\right], \\
  D_{\beta^{\ast}\beta}^c &= D_{\beta\beta^{\ast}}^c = \frac{D\gamma_b}{2},
    \label{eq:Dvan}
\end{align}
with all other elements vanishing.  The noise strength parameter $D$ depends on the environmental temperature, with $D = 1$ at zero temperature (quantum fluctuations persist even at absolute zero).  
For brevity, we can also define
\begin{align}
    \langle\xi^{\ast}_{\mu}(t)\xi_{\nu}(t')\rangle=D_{\mu\nu}\delta(t-t'),\quad \langle\xi_{\mu}(t)\xi_{\nu}(t')\rangle=0,
    \label{eq:classicalnoise2}
\end{align}
where
\begin{align}
  D_{\alpha\alpha}&= D\left[\frac{\gamma_a}{2} + \Gamma\left(2\langle|\alpha|^2\rangle - 1\right)\right], \label{eq:Dvan1}\\
  D_{\beta\beta} &= \frac{D\gamma_b}{2}.
    \label{eq:Dvan2}
\end{align}

\section{Nonlinear EP sensing in two-cavity model}

In this section, we elaborate on the nonlinear EP sensing scheme for the two-cavity model, and derive its Petermann factor using the standard approach (i.e., without accounting for the EP-shifting and other effects described in the main text).  We start with the equations of motion
\begin{align}
    \frac{d}{dt}\left[\begin{array}{c}
        \alpha\\
        \beta
    \end{array}\right]=-i\left[\begin{array}{cc}
        \omega_a + \frac{i}{2}\gamma(|\alpha|^2) & g\\
        g & \omega_b - \frac{i}{2}\gamma_b
    \end{array}\right]\left[\begin{array}{c}
    \alpha\\
    \beta
    \end{array}\right]+\left[\begin{array}{c}
        \xi_{\alpha}\\
        \xi_{\beta}
    \end{array}\right],
    \label{eq:classicallangevineq}
\end{align}
where the noise terms $\xi_{\alpha,\beta}$ are assumed to be independent of each other, have zero mean, and satisfy
\begin{align}
  \langle\xi^{\ast}_{\mu}(t)\xi_{\nu}(t^{\prime})\rangle
  &=D_{\mu\nu}\delta(t-t^{\prime}), \\
  \langle\xi_{\mu}(t)\xi_{\nu}(t^{\prime})\rangle &= 0.
\end{align}
The only non-zero elements of the diffusion matrix are the diagonal terms $D_{\alpha\alpha}$ and $D_{\beta\beta}$. The detailed form of $D_{\mu\nu}$ depends on the specific model. For van der Pol oscillators, $D_{\mu\nu}$ is given by Eqs.~\eqref{eq:Dvan1}-\eqref{eq:Dvan2}.

Consider the noise-free case, $\xi_{\alpha,\beta}=0$.  To determine the steady state, we assume the fields take the form $\alpha=\alpha_0 e^{-i\omega t}$ and $\beta=\beta_0 e^{-i\omega t}$, where $\alpha_0$ and $\beta_0$ are the steady-state amplitudes. Equation \eqref{eq:classicallangevineq} becomes
\begin{align}
\left[\begin{array}{cc}
        \omega_a + \frac{i}{2}\gamma(|\alpha_0|^2)-\omega & g\\
        g & \omega_b - \frac{i}{2}\gamma_b-\omega
    \end{array}\right]\left[\begin{array}{c}
    \alpha_0\\
    \beta_0
    \end{array}\right]=0.
    \label{eq:classicaleq}
\end{align}
To have a non-zero solution, we require the determinant of the matrix on the left-hand side to be zero. This yields the quadratic equation
\begin{align}
    \omega ^2 + \omega  \left[\frac{i \gamma_b-i \gamma(|\alpha_0|^2)}{2}-\omega_a-\omega_b\right]+\frac{\gamma(|\alpha_0|^2) \gamma_b}{4}+\frac{i \gamma(|\alpha_0|^2) \omega_b}{2}-\frac{i \gamma_b \omega_a}{2}-g^2+\omega_a \omega_b=0.
\end{align}
To solve this complex equation, we separate the real and imaginary parts:
\begin{align}
    \omega ^2- \omega  \left(\omega_a+\omega_b\right)+\frac{\gamma(|\alpha_0|^2) \gamma_b}{4}-g^2+\omega_a \omega_b&=0, \\
     \omega\left[\gamma_b-\gamma(|\alpha_0|^2)\right]+\gamma(|\alpha_0|^2) \omega_b-\gamma_b \omega_a&=0.
\end{align}
Eliminating $\gamma(|\alpha_0|^2)$ from the equations, we arrive at a cubic equation for $\omega$,
\begin{align}
    4(\omega-\omega_a)(\omega-\omega_b)^2 + (\omega-\omega_a)\gamma_b^2 - 4(\omega-\omega_b)g^2 = 0.
    \label{eq:polyomega}
\end{align}
After solving for $\omega$, we can express the nonlinear gain as
\begin{align}
    \gamma(|\alpha_0|^2)=\frac{4 \left[g^2-(\omega -\omega_a) (\omega -\omega_b)\right]}{\gamma_b}.
    \label{eq:gammaaeff}
\end{align}

Equation~\eqref{eq:polyomega} has a triple root at $\omega_a=\omega_b$, $\gamma_b=2g$. Along the parametric direction corresponding to the detuning $\Delta\omega=\omega_a-\omega_b$, the steady-state frequency varies to leading order as
\begin{align}
\omega=\omega_b+g|\eta|^{1/3}\text{sgn}(\eta),
\end{align} 
where $\eta\equiv\Delta\omega/g$. The nonlinear system thus behaves like a third-order EP.

We can use Eq.~\eqref{eq:gammaaeff} to get the nonlinear gain, $\gamma(|\alpha_0|^2)\approx2g(1-|\eta|^{2/3})$. To the leading order in $\eta$, the instantaneous Hamiltonian then takes the form
\begin{align}
    \matr{H}_0\approx\left[\begin{array}{cc}
        \omega_a+ig\left(1-|\eta|^{2/3}\right) & g\\
        g & \omega_b-ig
    \end{array}\right]\approx\left[\begin{array}{cc}
        \omega_b+ig\left(1-|\eta|^{2/3}\right) & g\\
        g & \omega_b-ig
    \end{array}\right].
\end{align}
The power-normalized right eigenstates of the above Hamiltonian are
\begin{align}
    \ket{\psi_{\pm}^R}\approx \frac{1}{(2+|\eta|^{2/3})^{1/2}}\left[
        \begin{array}{c}
          i\pm|\eta|^{1/3}\\
          1
            \end{array}
        \right].
\end{align}
Hence, the Petermann factor is given by
\begin{align}
    K=\frac{1}{|\braket{\psi_{\pm}^L}{\psi_{\pm}^R}|^2}\approx|\eta|^{-2/3}.
\end{align}

In the main text, we have discussed the performance of the nonlinear EP sensor along the detuning direction ($\Delta\omega$). Here, let us briefly discuss its performance along the loss rate direction ($\gamma_b$).  For small changes of loss rate $\Delta\gamma_b\equiv2g-\gamma_b$, the steady-state frequency changes as 
\begin{align}
    \begin{cases}
     \omega \approx \omega_b\pm g\sqrt{\delta}, & \text{if } \delta>0, \\
     \omega = \omega_b, & \text{otherwise},
   \end{cases}
\end{align} 
where $\delta\equiv\Delta\gamma_b/g$. This behaves like a second-order EP. We focus on the case of $\delta>0$. The nonlinear gain is obtained from Eq.~\eqref{eq:gammaaeff}, which is given by $\gamma(|\alpha_0|^2)\approx2g(1-\frac{\delta}{2})$. Hence, the instantaneous Hamiltonian to leading order in $\delta$ is
\begin{align}
    \matr{H}_0 \approx \left[\begin{array}{cc}
        \omega_b+ig\left(1-\frac{\delta}{2}\right) & g\\
        g & \omega_b-ig\left(1-\frac{\delta}{2}\right)
    \end{array}\right].
\end{align}
The right eigenstates are given by
\begin{align}
  \ket{\psi_{\pm}^R}\approx\frac{1}{\sqrt{\delta+2}}\left[
    \begin{array}{c}
      i\pm\sqrt{\delta}\\
      1
        \end{array}
    \right].
\end{align}
Thus, the Petermann factor is
\begin{align}
  K=\frac{1}{|\braket{\psi_{\pm}^L}{\psi_{\pm}^R}|^2}\approx \delta^{-1}.
\end{align}
The frequency uncertainty scales as $\sigma_{\omega}\propto K^{1/2}\propto\delta^{-1/2}$. Since the susceptibility follows $|\partial\omega/\partial\Delta\gamma_b|\propto\delta^{-1/2}$, which is at the same rate as the uncertainty $\sigma_{\omega}$, there is no SNR enhancement for nonlinear EP sensing along the $\gamma_b$ direction, even in the noise-free case.  Similar to the procedure in the main text, we can numerically simulate the time evolution of the noisy system and extract the dominant peak frequencies, uncertainties, and SNRs from the numerical power spectra. The van der Pol oscillator is used as the $a$ cavity. The results are shown in Fig.~\ref{fig:error_analysis_gammab_app}.  Near the EP, the uncertainty is significantly enhanced, and the susceptibility is degraded, leading to worse SNR.

\begin{figure}[t]
  \centering
  \includegraphics[width=0.5\linewidth]{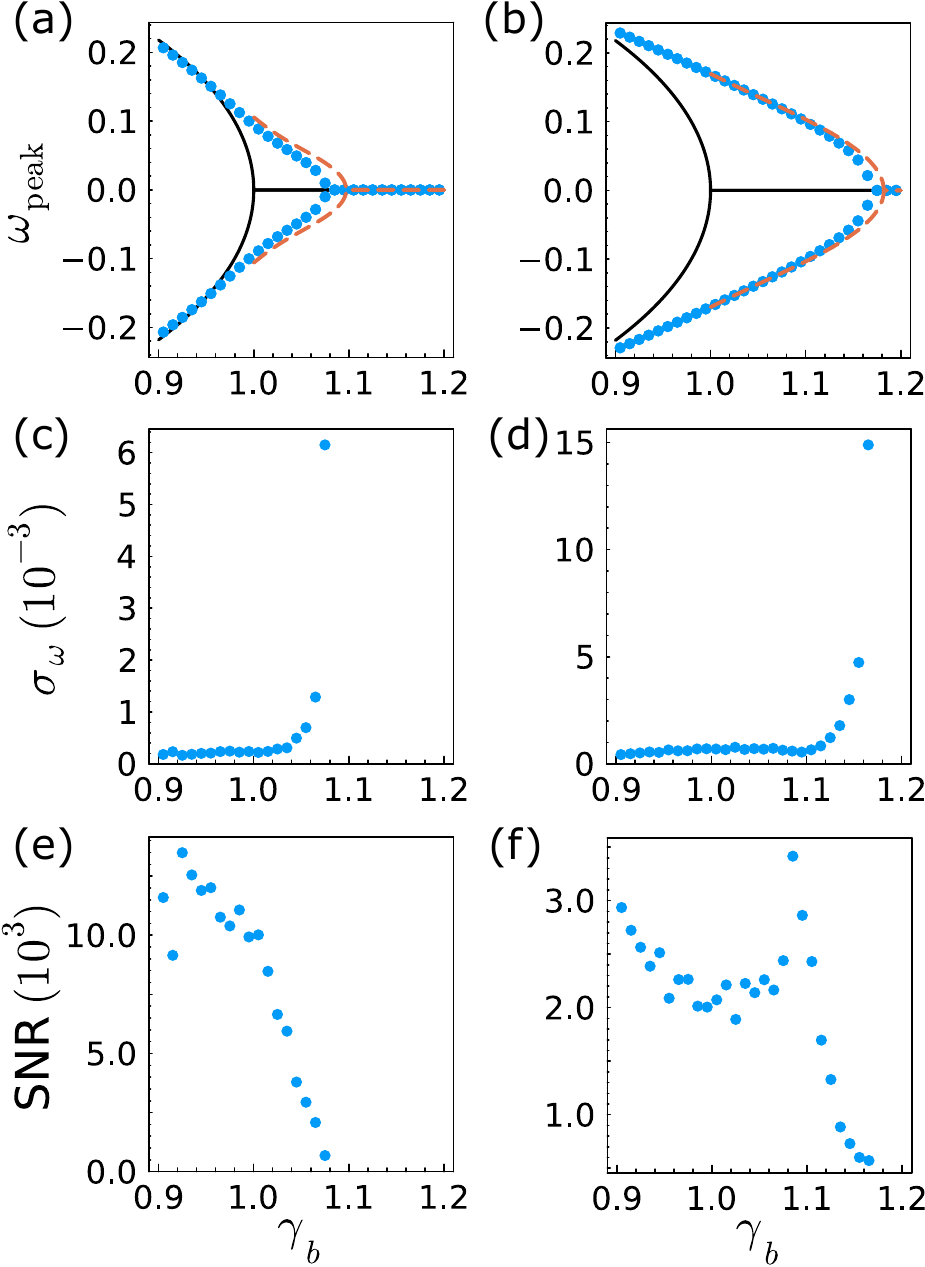}
  \caption{Effect of noise on nonlinear EP sensing along the loss rate ($\gamma_b$) direction. (a)-(b) Peak frequency $\omega_\text{peak}$ versus the loss rate $\gamma_b$ at different noise strength $D = 0.1$ and $D=1$. The blue dots denote the peak frequencies extracted from numerical simulations. The orange dashed lines represent the eigenfrequencies obtained from effective theory. The solid black lines represent the steady-state frequencies of noise-free system. (c)-(d) Uncertainties of peak frequency. (e)-(f) SNRs of the sensing. Parameters used are $\gamma_a=2$, $g=0.5$, $\Gamma=0.01$, $\omega_a=\omega_b=0$.}
  \label{fig:error_analysis_gammab_app}
\end{figure}

\section{Noise-adjusted effective theory}

Here, we derive a theoretical description of how the nonlinear two-cavity model's parameters are effectively adjusted by noise, giving rise to the shifted EP described in the main text.  Let us take the parameterization
\begin{align}
  \alpha &= Ae^{-i\phi_a}, \\
  \beta &= Be^{-i\phi_b}.
\end{align}
Equation~\eqref{eq:classicallangevineq} then becomes
\begin{align}
    \dot{A}&=\frac{1}{2}\gamma(A^2)A +gB\sin{\varphi}+\xi_A, \label{Adot} \\
    \dot{B}&=-\frac{\gamma_b}{2}B -gA\sin{\varphi}+\xi_B, \label{Bdot} \\
    \dot{\varphi}
    &= \Delta\omega - g\left(\frac{A}{B}-\frac{B}{A}\right) \cos{\varphi}
    +\xi_{\varphi},
    \label{eq:amplitudephase}
\end{align}
where $\varphi\equiv\phi_a-\phi_b$ is the relative phase between the two resonators.  The noise terms satisfy $\langle\xi_{\mu}(t)\xi_{\nu}(t^{\prime})\rangle=D_{\mu\nu}\delta(t-t^{\prime})$ with nonvanishing diffusion matrix elements
\begin{align}
 D_{AA}&=\frac{D_{\alpha\alpha}}{2},~D_{BB}=\frac{D_{\beta\beta}}{2},~D_{\varphi\varphi}=\frac{D_{\alpha\alpha}}{2\langle A^2\rangle}+\frac{D_{\beta\beta}}{2\langle B^2\rangle}.
    \label{eq:Ddiffusion}
\end{align}
To simplify matters, we expand the dynamical variables as the sum of their steady-state mean values and fluctuations, $S=S_0+\delta S$. The steady-state mean values satisfy
\begin{align}
    \left\{\frac{\gamma(A_0^2)}{2}+\left[\frac{3}{2}\gamma^{\prime}(A_0^2)+A_0^2\gamma^{\prime\prime}(A_0^2)\right]\langle\delta A^2\rangle\right\}A_0+gB_0\sin{\varphi_0}\left(1-\frac{1}{2}\langle\delta\varphi^2\rangle\right) &= 0,
    \label{eq:steadystateamplitude1} \\
    -\frac{\gamma_b}{2}B_0 -gA_0\sin{\varphi_0}\left(1-\frac{1}{2}\langle\delta\varphi^2\rangle\right)  &= 0,
    \label{eq:steadystateamplitude2} \\
    \Delta\omega - g\left(\frac{A_0}{B_0}-\frac{B_0}{A_0}\right)\cos{\varphi_0}\left(1-\frac{1}{2}\langle\delta\varphi^2\rangle\right) &= 0.
    \label{eq:steadystateamplitude3} 
\end{align}
Here we have kept the second-order terms of fluctuations in order to include the effect of the noise to the steady state, and assume the correlations between $\delta A$, $\delta B$ and $\delta\varphi$ are zero. The fluctuations satisfy
\begin{align}
    \delta\dot{A}&\approx\left[\frac{1}{2}\gamma(A_0^2)+\gamma^{\prime}(A_0^2)A_0^2\right]\delta A + g\sin{\varphi_0}\delta B + gB_0\cos{\varphi_0}\delta\varphi +\xi_A,\label{eq:dotA}\\
    \delta\dot{B}&\approx-\frac{\gamma_b}{2}\delta B - g\sin{\varphi_0}\delta A - gA_0\cos{\varphi_0}\delta\varphi +\xi_B,\label{eq:dotB}\\
    \delta\dot{\varphi}&\approx g\left(\frac{A_0}{B_0}-\frac{B_0}{A_0}\right)\sin{\varphi_0}\delta\varphi-g\left(\frac{1}{B_0}+\frac{B_0}{A_0^2}\right)\cos{\varphi_0}\delta A+g\left(\frac{1}{A_0}+\frac{A_0}{B_0^2}\right)\cos{\varphi_0}\delta B+\xi_{\varphi}\label{eq:dotphi}.
\end{align}

We aim to self-consistently solve Eqs.~\eqref{eq:steadystateamplitude1}-\eqref{eq:dotphi} to obtain both the steady state values (e.g., $A_0$) and the mean squared fluctuations (e.g., $\langle\delta A^2\rangle$).  From this, we will be able to determine the effective gain experienced by the $a$ cavity, which (according to a standard Taylor expansion) is given by
\begin{align}
    \gamma_{a,\text{eff}}=\langle\gamma(A^2)\rangle\approx\gamma(A_0^2)+\left[\gamma^{\prime}(A_0^2)+2A_0^2\gamma^{\prime\prime}(A_0^2)\right]\langle\delta A^2\rangle.
    \label{eq:gammaaeff2}
\end{align}
Then we can insert this into the effective Hamiltonian
\begin{align}
    \matr{H} = \left[
        \begin{array}{cc}
            \omega_a+\frac{i}{2}\gamma_{a,\text{eff}} & g\\
            g & \omega_b-\frac{i}{2}\gamma_b
        \end{array}
    \right],
\end{align} 
whose eigenvalues are the adjusted eigenfrequencies of the system.

We first consider the case of $\Delta\omega=0$.  We find that $\varphi_0=3\pi/2$, so then Eqs.~\eqref{eq:steadystateamplitude1}, \eqref{eq:steadystateamplitude2} and \eqref{eq:gammaaeff2} combine and simplify to give
\begin{align}
    \gamma_{a,\text{eff}}(\Delta\omega=0) \approx \frac{4g^2}{\gamma_b}\,
    \left(1-\langle\delta\varphi^2\rangle|_{\Delta\omega=0}\right)-2\gamma^{\prime}(A_0^2)\langle\delta A^2\rangle|_{\Delta\omega=0},
    \label{eq:gammaaeff0}
\end{align}
where $4g^2/\gamma_b$ is the effective gain in the noise-free case.  This manifestly shows that the effective gain is modified by the phase and amplitude fluctuations.

For $\Delta\omega\neq 0$, we start by ignoring the amplitude fluctuation terms, and focus first on phase fluctuations, whose dynamics satisfy
\begin{align}
    \delta\dot{\varphi}\approx g\left(\frac{A_0}{B_0}-\frac{B_0}{A_0}\right)\sin{\varphi_0}\delta\varphi+\xi_{\varphi}.
\end{align}
This is a Brownian oscillator, so a standard derivation with the aid of \eqref{eq:Ddiffusion} yields
\begin{align}
    \langle\delta\varphi^2\rangle=-\frac{D_{\varphi\varphi}}{2g\left(\frac{A_0}{B_0}-\frac{B_0}{A_0}\right)\sin{\varphi_0}}.
    \label{eq:phasevariance}
\end{align}
Combing Eqs.~\eqref{eq:steadystateamplitude1}-\eqref{eq:steadystateamplitude3},~\eqref{eq:gammaaeff2} and \eqref{eq:phasevariance} and ignoring amplitude fluctuations, we obtain
\begin{align}
    \frac{\Delta\omega^2}{g^2\left(\frac{\gamma_b}{\gamma_{a,\text{eff}}}+\frac{\gamma_{a,\text{eff}}}{\gamma_b}-2\right)}+\frac{\gamma_b\gamma_{a,\text{eff}}}{4g^2}=1-\frac{2D_{\varphi\varphi}}{D_{\varphi\varphi}+2\gamma_b-2\gamma_{a,\text{eff}}}.
\end{align}
In the limit $\Delta\omega/g\ll \sqrt{D_{\varphi\varphi}/g}\ll 1$, we keep terms up the leading order of $\Delta\omega$ and the first order of $D_{\varphi\varphi}$, and find
\begin{align}
    \gamma_{a,\text{eff}}\approx \gamma_{a,\text{eff}}(\Delta\omega=0)+\chi\Delta\omega^2,
    \label{eq:changeofgamma}
  \end{align}
where
\begin{align}
  \chi\approx-\frac{16g^2\gamma_b}{(\gamma_b^2-4g^2)^2}+\frac{16D_{\varphi\varphi}g^2\gamma_b^2(8g^2+\gamma_b^2)}{(\gamma_b^2-4g^2)^4}.
\end{align}
In the derivation of Eq.~\eqref{eq:changeofgamma} we have ignored the amplitude fluctuations. The change of amplitude fluctuations is of order $D\chi\Delta\omega^2$, which is a higher-order term.
Hence, the effective gain changes as $|\Delta\omega|^2$ near the EP, instead of $|\Delta\omega|^{2/3}$ in the noise-free case. This means the EP no longer behaves like a third-order EP.

Next, we can use the effective gain formula to find the shifted EP. At the shifted EP, we must have
\begin{align}
    \gamma_{a,\text{eff}}(\Delta\omega=0)+\gamma_b=4g.
\end{align}
Plugging Eq.~\eqref{eq:gammaaeff0} into the above equation, we obtain
\begin{align}
    \gamma_b^c \approx 2g \left(1 + \sqrt{\langle\delta\varphi^2\rangle+\frac{\gamma^{\prime}(A_0^2)\langle\delta A^2\rangle}{g}}+\frac{\gamma^{\prime}(A_0^2)\langle\delta A^2\rangle}{2g}\right)\Bigg|_{\Delta\omega=0},
    \label{eq:gammacritical}
\end{align}
Combining Eqs.~\eqref{eq:gammaaeff0},~\eqref{eq:changeofgamma} and \eqref{eq:gammacritical}, and keeping leading order terms of fluctuations, we find that the effective gain near the shifted EP is approximately given by
\begin{align}
    \gamma_{a,\text{eff}} \approx 2g\left(1-\sqrt{\langle\delta\varphi^2\rangle+\frac{\gamma^{\prime}(A_0^2)\langle\delta A^2\rangle}{g}}-\frac{\gamma^{\prime}(A_0^2)\langle\delta A^2\rangle}{2g}-\langle\delta\varphi^2\rangle\right)\Bigg|_{\Delta\omega=0}+\chi\Delta\omega^2.
\end{align}

\section{Noise amplification near the EP}

In this section, we derive the noise amplification near the shifted EP. Working near the shifted EP, let us take the expansion
\begin{align}
  \alpha &= A_0e^{-i\omega_0 t}+\delta\alpha \\
  \beta &= B_0e^{-i\omega_0 t+i\varphi_0}+\delta\beta.
\end{align}
Plugging this into Eq.~\eqref{eq:classicallangevineq} gives
\begin{align}
    \delta\dot{\alpha}&=\left[-i\tilde{\omega}_a+\frac{1}{2}\gamma(A_0^2)+\frac{1}{2}A_0^2\gamma^{\prime}(A_0^2)\right]\delta\alpha+\frac{1}{2}A_0^2\gamma^{\prime}(A_0^2)\delta\alpha^{\ast}-ig\delta\beta+\xi_{\alpha},
    \label{eq:fluctuationdynamics01}\\
    \delta\dot{\beta}&=\left(-i\tilde{\omega}_b-\frac{\gamma_b}{2}\right)\delta\beta-ig\delta\alpha+\xi_{\beta},
    \label{eq:fluctuationdynamics02}
\end{align}
where $\omega_0$ is the steady-state oscillation frequency of the mean field. We have transformed into a rotating frame of frequency $\omega_0$, such that $\tilde{\omega}_{a,b}=\omega_{a,b}-\omega_0$ is the co-rotating cavity frequency. The coefficient of $\delta\alpha$ in Eq.~\eqref{eq:fluctuationdynamics01} can be expressed in terms of $\gamma_{a,\text{eff}}=\langle\gamma(|\alpha|^2)\rangle$ as
\begin{align}
    \delta\dot{\alpha}&=\left(-i\tilde{\omega}_a+\frac{\gamma_{a,\text{eff}}}{2}+\frac{1}{2}A_0^2\gamma^{\prime}(A_0^2)\right)\delta\alpha+\frac{1}{2}A_0^2\gamma^{\prime}(A_0^2)\delta\alpha^{\ast}-ig\delta\beta+\xi_{\alpha}+\xi_{h},
    \label{eq:fluctuationdynamics1}\\
    \delta\dot{\beta}&=\left(-i\tilde{\omega}_b-\frac{\gamma_b}{2}\right)\delta\beta-ig\delta\alpha+\xi_{\beta},
    \label{eq:fluctuationdynamics2}
\end{align}
where $\xi_h$ is a high-order fluctuation term. Note that the dynamics of $\delta\alpha$ involve its conjugate $\delta\alpha^{\ast}$. We can study the fluctuation dynamics in the basis $\ket{\Psi}=[\delta\alpha,\delta\beta,\delta\alpha^{\ast},\delta\beta^{\ast}]^T$. Equations \eqref{eq:fluctuationdynamics1}-\eqref{eq:fluctuationdynamics2} then become
\begin{align}
    i\ket{\dot{\Psi}}&=\matr{H}_B\ket{\Psi}+\ket{\xi},
    \label{eq:bdgdynamics}
\end{align}
where
\begin{align}
    \matr{H}_B&=\left[\begin{array}{cc}
        \matr{H}_0+\matr{V}_0 & \matr{V}_0\\
        -\matr{V}_0^{\ast} & -\matr{H}_0^{\ast}-\matr{V}_0^{\ast},\\
      \end{array}\right],     
      \label{eq:bdg}\\[5pt]
    \matr{H}_0&=\left[
          \begin{array}{cc}
              \tilde{\omega}_a+\frac{i}{2}\gamma_{a,\text{eff}} & g\\
              g & \tilde{\omega}_b-\frac{i}{2}\gamma_b
          \end{array}
      \right],
      \label{eq:H01}\\[5pt]
    \matr{V}_0&=\left[\begin{array}{cc}
          \frac{i}{2}A_0^2\gamma^{\prime}(A_0^2)& 0\\
        0 & 0
        \end{array}\right],
        \label{eq:v0dv}
        \\[5pt]
    \ket{\xi}&=[\xi_{\alpha}+\xi_h,\xi_{\beta},\xi_{\alpha}^{\ast}+\xi_h^{\ast},\xi_{\beta}^{\ast}]^T.
\end{align}
Next, we neglect the high-order fluctuation term $\xi_h$, so that $\ket{\xi}\approx[\xi_{\alpha},\xi_{\beta},\xi_{\alpha}^{\ast},\xi_{\beta}^{\ast}]^T$. The correlation of $\ket{\xi}$ satisfies
\begin{align}
  \langle\ket{\xi(t)}\bra{\xi(t^{\prime})}\rangle
  &= \matr{D}_{\xi\xi}\delta(t-t^{\prime}),\\ \matr{D}_{\xi\xi}
  &\approx
  \begin{pmatrix}
    D_{\alpha\alpha} & & & \\ & D_{\beta\beta} & &\\ & & D_{\alpha\alpha}&
    \\&&&D_{\beta\beta}
  \end{pmatrix}.
\end{align}

For $\Delta\omega=0$ and $\gamma_b=\gamma_b^c$, the eigenvalues of the Bogoliubov-de Gennes Hamiltonian $\matr{H}_B$ are
\begin{align}
    E_{1,2}=E_0,~E_{3,4}=E_0+\frac{i}{2}A_0^2\gamma^{\prime}(A_0^2)\pm\sqrt{-A_0^2\gamma^{\prime}(A_0^2)\left[1+\frac{1}{4}A_0^2\gamma^{\prime}(A_0^2)\right]}
\end{align}
where 
\begin{align}
    E_0=\tilde{\omega}_a-i\gamma_0,\quad \gamma_0\equiv\frac{1}{4}[\gamma_b-\gamma_{a,\text{eff}}(\Delta\omega=0)].
\end{align}
There is a pair of degenerate eigenvalues $E_{1,2}$ whose corresponding right eigenstates are
\begin{align}
  \ket{\psi^R_{1,2}} =
  \frac{1}{2}
  \begin{pmatrix}
    i\\1\\-i\\1
  \end{pmatrix}.
\end{align}
This means $\matr{H}_B$ has a EP at $\Delta\omega=0$, $\gamma_b=\gamma_b^c$, which coincides with the shifted EP of the Hamiltonian $\matr{H}_0$.

For $\Delta\omega\neq 0$, the degenerate eigenvalues and corresponding eigenstates undergo splitting. To leading order in $\eta\equiv\Delta\omega/g$, the eigenvalues and eigenstates are approximated as follows:
\begin{align}
  E_{\pm} &\approx E_0\pm k_0\eta,\\
  k_0 &= \sqrt{\frac{2g+(\chi g+2)A_0^2\gamma^{\prime}(A_0^2)}{-2A_0^2\gamma^{\prime}(A_0^2)}},\\
  \ket{\psi^{R}_{\pm}} &\approx
  \frac{1}{2}
  \begin{pmatrix}
    i+\left(-1-\frac{2}{gA_0^2\gamma^{\prime}(A_0^2)}\pm k_0\right)\eta \\
    1+i\left(2+\frac{2}{gA_0^2\gamma^{\prime}(A_0^2)}\right)\eta \\
    -i+(1\mp k_0)\eta \\
    1
  \end{pmatrix}.
  \label{eq:eigenstates}
\end{align}

To analyze the noise near the shifted EP, we need to solve Eq.~\eqref{eq:bdgdynamics}. In the frequency domain,
\begin{align}
    \ket{\Psi(\omega)}&=\matr{G}(\omega)\ket{\xi(\omega)},\\
    \matr{G}(\omega)&=(\omega\matr{I}-\matr{H}_B)^{-1}.
\end{align}
The Green's function $\matr{G}(\omega)$ can be expressed in terms of the eigenvalues and eigenstates of the Hamiltonian $\matr{H}_B$ as
\begin{align}
    \matr{G}(\omega)=\sum_i\frac{1}{\omega-E_i}\frac{\ket{\psi^R_i}\bra{\psi^L_i}}{\braket{\psi^L_i}{\psi^R_i}}.
\end{align}
The dominant states are those that are close to the threshold, i.e., $\text{Im}(E_i)$ are close to zero. Near the shifted EP,
\begin{equation}
  \text{Im}(E_{\pm})=-\gamma_0\approx -4g \sqrt{\langle\delta\varphi^2\rangle+\frac{\gamma^{\prime}(A_0^2)\langle\delta A^2\rangle}{g}}.
\end{equation}
Hence, we only consider the states $\ket{\psi^R_{\pm}}$. The Green's function $\matr{G}(\omega)$ can then be approximated as
\begin{align}
    \matr{G}(\omega)\approx \frac{1}{\omega-\tilde{\omega}_a+i\gamma_0-k_0\eta}\frac{\ket{\psi^R_+}\bra{\psi^L_+}}{\braket{\psi^L_+}{\psi^R_+}}+\frac{1}{\omega-\tilde{\omega}_a+i\gamma_0+k_0\eta}\frac{\ket{\psi^R_-}\bra{\psi^L_-}}{\braket{\psi^L_-}{\psi^R_-}}.
    \label{eq:approxgreen}
\end{align}
In the time domain,
\begin{align}
    \ket{\Psi(t)}=\int^tdt^{\prime}\int \frac{d\omega}{2\pi}\matr{G}(\omega)e^{-i\omega(t-t^{\prime})}\ket{\xi(t^{\prime})}.
    \label{eq:timedomainevo}
\end{align}
Plugging Eq. \eqref{eq:approxgreen} into Eq. \eqref{eq:timedomainevo} and doing the contour integral,
we get
\begin{align}
    \ket{\Psi(t)}\approx-i\int^tdt^{\prime}\left[\frac{\ket{\psi^R_+}\braket{\psi^L_+}{\xi(t^{\prime})}}{\braket{\psi^L_+}{\psi^R_+}}e^{-(\gamma_0+i\tilde{\omega}_a+ik_0\eta)(t-t^{\prime})}+\frac{\ket{\psi^R_-}\braket{\psi^L_-}{\xi(t^{\prime})}}{\braket{\psi^L_-}{\psi^R_-}}e^{-(\gamma_0+i\tilde{\omega}_a-ik_0\eta)(t-t^{\prime})}\right],
\end{align}
The fluctuation $\delta\alpha$ is then given by
\begin{align}
    \delta\alpha(t)\approx-i\int^tdt^{\prime}\left[\frac{\psi^R_{+,1}\braket{\psi^L_+}{\xi(t^{\prime})}}{\braket{\psi^L_+}{\psi^R_+}}e^{-(\gamma_0+i\tilde{\omega}_a+ik_0\eta)(t-t^{\prime})}+\frac{\psi^R_{-,1}\braket{\psi^L_-}{\xi(t^{\prime})}}{\braket{\psi^L_-}{\psi^R_-}}e^{-(\gamma_0+i\tilde{\omega}_a-ik_0\eta)(t-t^{\prime})}\right],
\end{align}
where $\psi^R_{\pm,1}$ represents the first component of the eigenstates. The correlation $\langle\delta\alpha^{\ast}(t_1)\delta\alpha(t_2)\rangle$ is given by the sum of four terms:
\begin{align}
    (\text{i}):
    &\int^{t_1}dt_1^{\prime}\int^{t_2}dt_2^{\prime}\langle\frac{|\psi^R_{+,1}|^2\braket{\psi^L_+}{\xi(t_2^{\prime})}\braket{\xi(t_1^{\prime})}{\psi^L_+}}{|\braket{\psi^L_+}{\psi^R_+}|^2}\rangle e^{-(\gamma_0+i\tilde{\omega}_a+ik_0\eta)(t_2-t_2^{\prime})-(\gamma_0-i\tilde{\omega}_a-ik_0\eta)(t_1-t_1^{\prime})}\nonumber\\
    =&\frac{|\psi^R_{+,1}|^2\bra{\psi^L_+}\matr{D}_{\xi\xi}\ket{\psi^L_+}}{|\langle\psi^L_+|\psi^R_+\rangle|^2}\int^{t_1}dt_1^{\prime}\int^{t_2}dt_2^{\prime}\delta(t_1^{\prime}-t_2^{\prime}) e^{-(\gamma_0+i\tilde{\omega}_a+ik_0\eta)(t_2-t_2^{\prime})-(\gamma_0-i\tilde{\omega}_a-ik_0\eta)(t_1-t_1^{\prime})}\nonumber\\
    =&\frac{|\psi^R_{+,1}|^2\bra{\psi^L_+}\matr{D}_{\xi\xi}\ket{\psi^L_+}}{2\gamma_0|\langle\psi^L_+|\psi^R_+\rangle|^2}e^{-\gamma_0|t_1-t_2|+i(\tilde{\omega}_a+k_0\eta)(t_1-t_2)},\\
    (\text{ii}):
    &\int^{t_1}dt_1^{\prime}\int^{t_2}dt_2^{\prime}\langle\frac{|\psi^R_{-,1}|^2\braket{\psi^L_-}{\xi(t_2^{\prime})}\braket{\xi(t_1^{\prime})}{\psi^L_-}}{|\braket{\psi^L_-}{\psi^R_-}|^2}\rangle e^{-(\gamma_0+i\tilde{\omega}_a-ik_0\eta)(t_2-t_2^{\prime})-(\gamma_0-i\tilde{\omega}_a+ik_0\eta)(t_1-t_1^{\prime})}\nonumber\\
    =&\frac{|\psi^R_{-,1}|^2\bra{\psi^L_-}\matr{D}_{\xi\xi}\ket{\psi^L_-}}{2\gamma_0|\braket{\psi^L_-}{\psi^R_-}|^2}e^{-\gamma_0|t_1-t_2|+i(\tilde{\omega}_a-k_0\eta)(t_1-t_2)},\\
    (\text{iii}):
    &\int^{t_1}dt_1^{\prime}\int^{t_2}dt_2^{\prime}\langle\frac{\psi^R_{+,1}\psi^{R\ast}_{-,1}\braket{\psi^L_+}{\xi(t_2^{\prime})}\braket{\xi(t_1^{\prime})}{\psi^L_-}}{\braket{\psi^L_+}{\psi^R_+}\braket{\psi^R_-}{\psi^L_-}}\rangle e^{-(\gamma_0+i\tilde{\omega}_a+ik_0\eta)(t_2-t_2^{\prime})-(\gamma_0-i\tilde{\omega}_a+ik_0\eta)(t_1-t_1^{\prime})}\nonumber\\
    =&\frac{\psi^R_{+,1}\psi^{R\ast}_{-,1}\bra{\psi^L_+}\matr{D}_{\xi\xi}\ket{\psi^L_-}}{\braket{\psi^L_+}{\psi^R_+}\braket{\psi^R_-}{\psi^L_-}}\int^{t_1}dt_1^{\prime}\int^{t_2}dt_2^{\prime}\delta(t_1^{\prime}-t_2^{\prime}) e^{-(\gamma_0+i\tilde{\omega}_a+ik_0\eta)(t_2-t_2^{\prime})-(\gamma_0-i\tilde{\omega}_a+ik_0\eta)(t_1-t_1^{\prime})}\nonumber\\
    =&\frac{\psi^R_{+,1}\psi^{R\ast}_{-,1}\bra{\psi^L_+}\matr{D}_{\xi\xi}\ket{\psi^L_-}}{2(\gamma_0+ik_0\eta)\braket{\psi^L_+}{\psi^R_+}\braket{\psi^R_-}{\psi^L_-}} e^{-(\gamma_0+ik_0\eta)|t_1-t_2|+i\tilde{\omega}_a(t_1-t_2)},\\
    (\text{iv}):
    &\int^{t_1}dt_1^{\prime}\int^{t_2}dt_2^{\prime}\langle\frac{\psi^R_{-,1}\psi^{R\ast}_{+,1}\braket{\psi^L_-}{\xi(t_2^{\prime})}\braket{\xi(t_1^{\prime})}{\psi^L_+}}{\braket{\psi^L_-}{\psi^R_-}\braket{\psi^R_+}{\psi^L_+}}\rangle e^{-(\gamma_0+i\tilde{\omega}_a-ik_0\eta)(t_2-t_2^{\prime})-(\gamma_0-i\tilde{\omega}_a-ik_0\eta)(t_1-t_1^{\prime})}\nonumber\\
    =&\frac{\psi^R_{-,1}\psi^{R\ast}_{+,1}\bra{\psi^L_-}\matr{D}_{\xi\xi}\ket{\psi^L_+}}{2(\gamma_0-ik_0\eta)\braket{\psi^L_-}{\psi^R_-}\braket{\psi^R_+}{\psi^L_+}} e^{-(\gamma_0-ik_0\eta)|t_1-t_2|+i\tilde{\omega}_a(t_1-t_2)}.
\end{align}
Using Eq.~\eqref{eq:eigenstates}, we have, to the leading order of $\eta$, 
\begin{align}
    &|\psi^R_{\pm,1}|^2\approx\psi^R_{\pm,1}\psi^{R\ast}_{\mp,1}\approx 1/4,\\
&K^{-1}\equiv|\braket{\psi^L_{\pm}}{\psi^R_{\pm}}|^2\approx\braket{\psi^L_{\pm}}{\psi^R_{\pm}}\braket{\psi^R_{\mp}}{\psi^L_{\mp}}\approx (k_0\eta)^{2},\\
    &\langle\psi^L_{\pm}|\matr{D}_{\xi\xi}|\psi^L_{\pm}\rangle\approx\langle\psi^L_{\pm}|\matr{D}_{\xi\xi}|\psi^L_{\mp}\rangle\approx (D_{\alpha\alpha}+D_{\beta\beta})/2.
\end{align}
The correlation $\langle\delta\alpha^{\ast}(t_1)\delta\alpha(t_2)\rangle$ is then simplified as
\begin{align}
    &\langle\delta\alpha^{\ast}(t_1)\delta\alpha(t_2)\rangle= (\text{i}) + (\text{ii}) + (\text{iii}) + (\text{iv}) \nonumber\\
    \approx&\frac{1}{8}(D_{\alpha\alpha}+D_{\beta\beta})Ke^{-\gamma_0|t_1-t_2|+i\tilde{\omega}_a(t_1-t_2)}\left[\frac{\cos{(k_0\eta|t_1-t_2|)}}{\gamma_0}+\frac{\gamma_0\cos{(k_0\eta|t_1-t_2|)}-k_0\eta\sin{(k_0\eta|t_1-t_2|)}}{\gamma_0^2+k_0^2\eta^2}\right].
\end{align}
The phase fluctuation can be obtained via
\begin{align}
    \delta\phi_a=-\frac{i(\delta\alpha^{\ast}-\delta\alpha)}{2A_0},
\end{align}
and the phase correlation is
\begin{align}
    \langle\delta\phi_a(t_1)\delta\phi_a(t_2)\rangle=\frac{\text{Re}[\langle\delta\alpha^{\ast}(t_1)\delta\alpha(t_2)\rangle]}{A_0^2}.
\end{align}
The phase uncertainty accumulated over time $\Delta t$ is
\begin{align}
    \langle(\delta\phi_a(t+\Delta t)-\delta\phi_a(t))^2\rangle=\frac{(D_{\alpha\alpha}+D_{\beta\beta})K}{2A_0^2}\Delta t+O(\Delta t^2).
\end{align}
Hence, the frequency uncertainty is
\begin{align}
    \sigma_{\omega}^2=\frac{\langle(\delta\phi_a(t+\Delta t)-\delta\phi_a(t))^2\rangle}{\Delta t^2}\approx\frac{(D_{\alpha\alpha}+D_{\beta\beta})K}{2A_0^2\Delta t}\approx\frac{(D_{\alpha\alpha}+D_{\beta\beta})}{2A_0^2k_0^2\Delta t}\frac{1}{\eta^2}.  
    \label{eq:sigmaanalytical}
\end{align}

\section{Curve fitting of numerical spectrum}

\begin{figure}[t]
  \centering
  \includegraphics[width=0.7\linewidth]{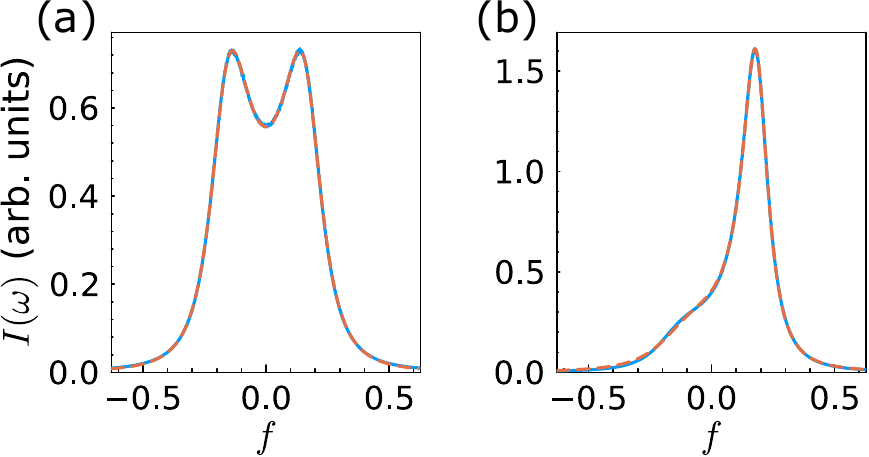}
  \caption{Curve fitting of the spectrum. (a) Symmetric spectrum with $\omega_a=\omega_b=0$. (b) Asymmetric spectrum with $\omega_a=0.03$ and $\omega_b=0$. In both (a) and (b), the blue solid lines show the spectra obtained from simulations, and the red dashed lines show the fitted curves based on Eqs. \eqref{eq:generalspectrumformula} and \eqref{eq:generalspectrumformula2}. The parameters used are $\gamma_a=2$, $\gamma_b=1$, $g=0.5$, $\Gamma=0.01$, $D=1$.} 
  \label{fig:curvefit}
\end{figure}

In this section, we describe the simulations and curve fitting used in the main text. For each simulation run, we use a fixed initial state $\alpha(0)=7$, $\beta(0)=-7i$ with an independent noise realization. Each run consists of a warm-up interval of $\Delta t_0=10^4$, followed by a data collection interval of $\Delta t=10^7$.  To balance the resolution and smoothness of the resulting power spectrum, we subdivide the latter time series into 1000 shorter intervals of $10^4$, calculate the spectra, and take the mean of the spectra.

To determine the peak frequency and frequency uncertainty from the numerical spectra, we use curve fitting. For sensing along the $\gamma_b$ direction, with $\Delta\omega=0$, we note that the spectrum is symmetric about the cavity frequency $\omega_a=\omega_b$, so without loss of generality we fit to the symmetric formula
\begin{align}
I(\omega) = \frac{p_4\omega^2+p_3}{p_2\omega^4 + p_1\omega^2 + p_0},
\label{eq:generalspectrumformula}
\end{align}
where $p_0, \dots, p_4$ are fitting parameters.  From this, we extract the peak frequency from the roots of the polynomial in the denominator. Fig.~\ref{fig:curvefit}(a) shows that the spectrum of $a$ cavity obtained from simulation (blue solid line) fits well with Eq.~\eqref{eq:generalspectrumformula} (red dashed line).

When sensing along the $\Delta\omega$ direction, the spectrum is asymmetric. In this case, we fit to the formula
\begin{align}
I(\omega)=\frac{p_7\omega^2+p_6\omega+p_5}{p_4\omega^4+p_3\omega^3+p_2\omega^2+p_1\omega+p_0}
\label{eq:generalspectrumformula2}
\end{align}
where $p_0, \dots, p_7$ are fitting parameters.  Again, we extract the peak frequencies from the roots of the polynomial in the denominator.  As shown in Fig.~\ref{fig:curvefit}(b), the numerical spectrum (blue solid line) is well fitted by Eq.~\eqref{eq:generalspectrumformula2} (red dashed line). The frequency uncertainty here is obtained by calculating the standard deviation of frequencies extracted from 100 independent simulation runs.

\section{Nonlinear EP sensing with a laser model}
\label{sec:lasermodel}

\begin{figure}[t]
  \centering
  \includegraphics[width=0.7\linewidth]{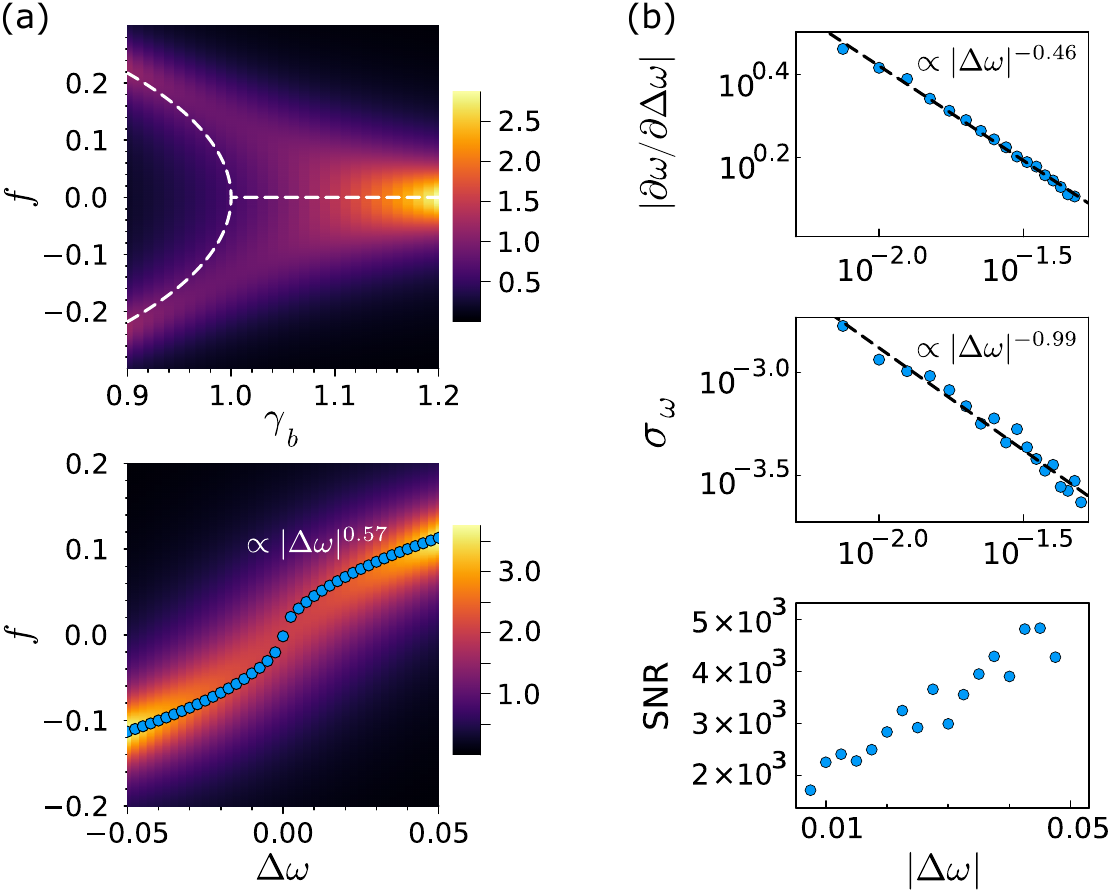}
  \caption{Nonlinear EP sensing based on laser model.  (a) Spectra as a function of the loss rate (top panel) and detuning (bottom panel). The white dashed line represents the steady-state frequency of the noise-free system. Blue dots indicate the dominant peak frequencies extracted from the spectra. In the top panel, we set $\omega_a = \omega_b = 0$, and in the bottom panel, $\gamma_b = 1.148$. (b) Susceptibility, frequency uncertainty, and SNR extracted from the numerical spectra near the shifted EP at $\gamma_b = 1.148$. Model parameters used are $\omega_a = 0$, $\gamma_a = 1$, $\Gamma = 3$, $r = 0.01$, $g = 0.5$, and $D = 1$.}
  \label{fig:lasingmodel}
\end{figure}

The analytical results in the main text were derived by assuming the $a$ resonator has a nonlinear, intensity-dependent gain.  We specifically modeled this as a van der Pol oscillator, obtaining good agreement with analytical predictions. Here, we present numerical results using an alternative nonlinearity that represents a coupled-cavity laser system.  This model is also the same one used in Ref.~\cite{bai2023nonlinearity}

The $c$-number Langevin equations for this model have the form
\begin{align}
\dot{\alpha} &= \left[-i\omega_a - \frac{\gamma_a}{2} + \frac{\Gamma}{2(1+r|\alpha|^2)}\right]\alpha - ig\beta + \xi_\alpha, \nonumber\\
\dot{\beta} &= \left(-i\omega_b - \frac{\gamma_b}{2}\right)\beta - ig\alpha + \xi_\beta,
\label{eq:lasingmodeleq}
\end{align}
where $\gamma_{a,b}$ is the linear loss rate of the $a$ and $b$ cavity, $\Gamma$ is the pumping rate, $r$ is the saturation coefficient, and $\xi_{\alpha,\beta}$ are noise fluctuations whose means are zero and their correlation functions can be determined by the corresponding Heisenberg-Langevin equation \cite{scully1997quantum}. For simplicity, we assume $\xi_{\beta}$ vanishes, and $\xi_{\alpha}$ obeys
\begin{align}
    \langle\xi_{\alpha}^{\ast}(t)\xi_{\alpha}(t^{\prime})\rangle=D\delta(t-t^{\prime}).
\end{align}
In the absence of noise, the system exhibits an EP at $\gamma_b=2g$ and $\Delta\omega\equiv\omega_a-\omega_b=0$. When noise is introduced, the spectra along the loss rate and detuning directions, obtained from numerical simulations, are shown in Fig.~\ref{fig:lasingmodel}(a). A shift in the EP is evident in the top panel, with the new EP located at $\gamma_b=1.148$. Near this shifted EP, the dominant peak frequency scales as $|\Delta\omega|^{0.57}$, as shown in the bottom panel, indicating a transition from a third-order-like EP to a second-order EP.

We also examine the susceptibility, frequency uncertainty, and SNR near the shifted EP, finding scaling behaviors similar to those observed in the van der Pol model and consistent with theoretical predictions. The results are shown in Fig.~\ref{fig:lasingmodel}(b).

\section{Effects of noise level on nonlinear EP sensing}

\begin{figure}[t]
    \centering
    \includegraphics[width=0.75\linewidth]{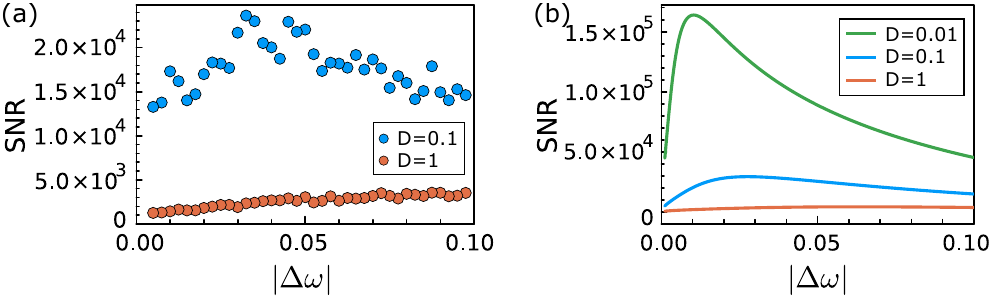}
    \caption{Effect of noise strength on SNR near the shifted EP in the van der Pol model. (a) SNR as a function of detuning, obtained from numerical simulations for noise strengths $D = 0.1$ and $D = 1$. The shifted EP occurs at $\gamma_b = 1.171$ for $D = 1$ and $\gamma_b = 1.09$ for $D = 0.1$. (b) SNR as a function of detuning, based on effective theory, for $D = 0.01$, $D = 0.1$, and $D = 1$. The shifted EP occurs at $\gamma_b = 1.182$ for $D = 1$, $\gamma_b = 1.097$ for $D = 0.1$, and $\gamma_b = 1.047$ for $D = 0.01$. The model parameters are $\gamma_a = 2$, $\Gamma = 0.01$, and $g = 0.5$.}
      \label{fig:noiselevel}
\end{figure}

As noted in the main text, the precise parametric dependence of the SNR depends on the details of the model, including the choice of noise level.  Here, we present some analytical numerical results to demonstrate this, for both the van der Pol oscillator model studied in the main text, and the laser model of Section~\ref{sec:lasermodel}.

Eq.~\eqref{eq:changeofgamma} shows that $\gamma_{a,\text{eff}}\propto \Delta\omega^2$ when $|\Delta\omega|/g\ll\sqrt{D_{\varphi\varphi}/g}$, whereas in the noise-free case, $\gamma_{a,\text{eff}}\propto |\Delta\omega|^{2/3}$. As a result, no SNR enhancement is observed for sufficiently small detuning in the presence of noise.  However, if the noise level is further reduced, the dependence becomes different at larger values of $|\Delta\omega|$.

In Fig.~\ref{fig:noiselevel}, we further present the SNR as a function of detuning for various noise levels, using both numerical simulations and effective theory. Regardless of the noise strength, the SNR decreases as $\Delta\omega$ is very close to the EP. However, lower noise levels narrow the region where the SNR starts to decline and increases the peak SNR value. Consequently, behaviors similar to the noise-free predications can still be observed in the regime $|\Delta\omega|/g\gtrapprox\sqrt{D_{\varphi\varphi}/g}$, corresponding to low noise level and moderately large detuning.

Similar behaviors are also found for the laser model. In Fig.~\ref{fig:nsrresult}(a)-(b), we show the spectra as a function of the detuning and loss rate near the noise-free EP for a low noise level ($D=0.01$) based on the laser model. While there is a shift in the EP, as evidenced by the peak splitting at the original EP [Fig.\ref{fig:nsrresult}(c)], the peak frequency agrees well with noise-free predications when the detuning is away from the EP, as shown in Fig. \ref{fig:nsrresult}(d). Additionally, we calculated the frequency uncertainties [Fig.\ref{fig:nsrresult}(e)] and the SNRs [Fig.\ref{fig:nsrresult}(f)]. As shown in Fig.\ref{fig:nsrresult}(f), when the noise is sufficiently low, the SNR initially increases, as expected by the noise-free prediction, and then declines, as predicted by our effective theory in the region $|\Delta\omega|/g\ll\sqrt{D_{\varphi\varphi}/g}$. 

\begin{figure}[t]
    \centering
    \includegraphics[width=0.7\linewidth]{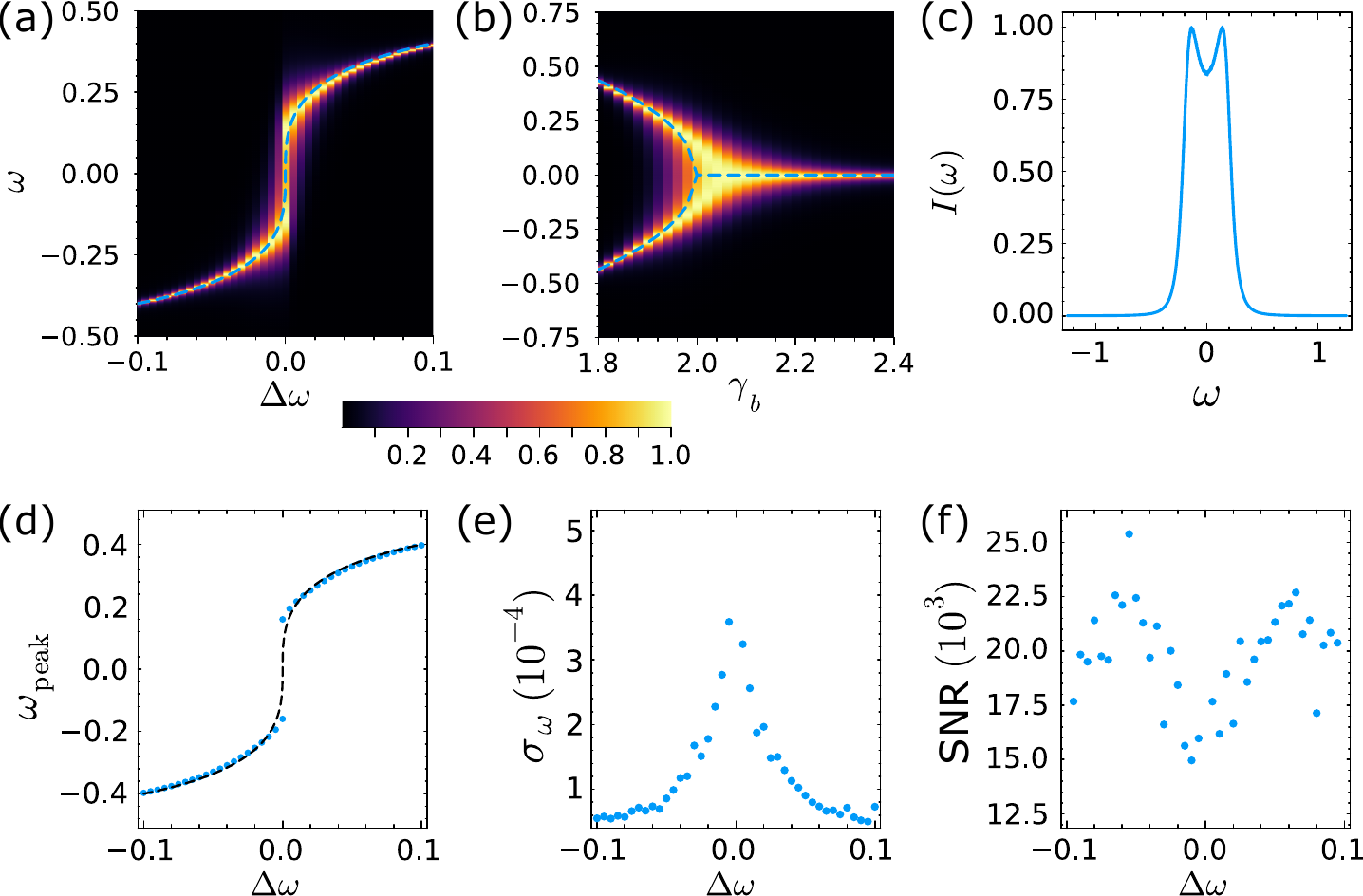}
    \caption{Nonlinear EP-sensing in low-noise regime of the laser model of Section~\ref{sec:lasermodel}. (a)-(b) Power spectra plotted against the detuning $\Delta\omega$ and loss rate $\gamma_b$ near the original EP. The blue dashes denote the steady-state frequency of the noise-free system. $\gamma_b=2$ in (a) and $\Delta\omega=0$ in (b). (c) Spectrum at the original EP. $\gamma_b=2$, $\Delta\omega=0$. (d)-(e) The mean and uncertainty of the peak frequency extracted from the power spectrum. (f) SNRs of the sensing scheme. In (d)-(f), $\gamma_b=2$. The model used is the same as Sec. \ref{sec:lasermodel}. The parameters are $\gamma_a=2$, $g=1$, $\Gamma=12$, $D=0.01$. }
      \label{fig:nsrresult}
\end{figure}

\section{EP sensing in a linear two-cavity system}

\begin{figure}[t]
    \centering
    \includegraphics[width=0.75\linewidth]{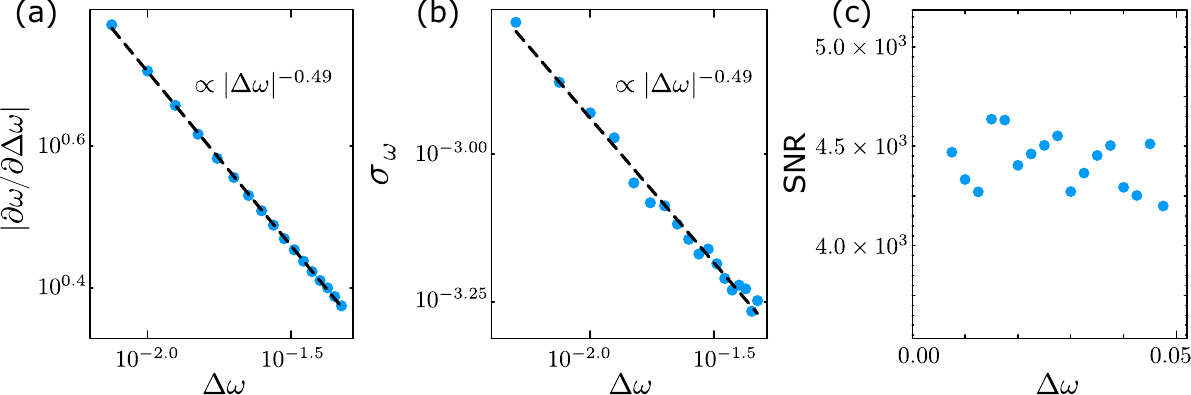}
    \caption{EP sensing in a linear two-cavity system. (a) - (c) Susceptibility, uncertainty and SNR as functions of detuning, obtained from curve fitting of simulated output spectra. The model parameters are $\gamma_a=1.3$, $\gamma_b=0.7$, $g=0.5$, $D=1.4$, $\alpha_{\textrm{in}}=1$.}
      \label{fig:linearep}
  \end{figure}

In this section, we present the results of EP sensing based on a linear two-cavity system. The equations of motion governing the system are:
\begin{align}
    \dot{\alpha}&=\left[-i\omega_a-\frac{\gamma_a}{2}\right]-ig\beta + \sqrt{\gamma_a}\alpha_{\textrm{in}}, \label{eq:linearep1}\\
    \dot{\beta}&=\left[-i\omega_b+\frac{\gamma_b}{2}\right]-ig\alpha + \sqrt{\gamma_b}\xi,
    \label{eq:linearep2}
\end{align}
where $\alpha_{\textrm{in}}$ is the input filed, $\xi$ is the noise and satisfies $\langle\xi^{\ast}(t)\xi(t^{\prime})\rangle=D\delta(t-t^{\prime})$. For simplicity, we have assumed there are no noise in the first cavity. We can solve Eqs.~\eqref{eq:linearep1}-\eqref{eq:linearep2} in the frequency domain
\begin{align}
    \alpha(\omega)=\frac{\sqrt{\gamma_a}\left(-\frac{\gamma_b}{2}-i\omega+i\omega_b\right)\alpha_{\textrm{in}}(\omega)-i\sqrt{\gamma_b}g\xi(\omega)}{g^2+\left(\frac{\gamma_a}{2}-i\omega+i\omega_a\right)\left(-\frac{\gamma_b}{2}-i\omega+i\omega_b\right)}.
\end{align}
The output field is determined using the input-output relation $\alpha_{\textrm{out}}(\omega)=\sqrt{\gamma_a}\alpha(\omega)-\alpha_{\textrm{in}}(\omega)$, and the output spectrum is defined as $S(\omega)=\textrm{Re}[\alpha_{\textrm{out}}(\omega)/\alpha_{\textrm{in}}(\omega)]$. 

In our simulation, the noise $\xi(\omega)$ is modeled as a complex random variable, $\mathcal{N}(0,Dd\omega/2)+i\mathcal{N}(0,Dd\omega/2)$, where $d\omega$ is the frequency resolution. We preform 1000 independent simulations, applying curve fitting to the output spectra to determine the mean peak frequencies, uncertainties and the SNR as before. The numerical results are shown in Fig. \ref{fig:linearep}. Both the susceptibility and uncertainty scale as $|\Delta\omega|^{-0.49}$, closely matching the scaling $|\Delta\omega|^{-1/2}$ predicted by linear theory. The amplification of susceptibility and uncertainty precisely cancel each other, resulting in no SNR enhancement.

\section{Nonlinear EP sensing in a three-cavity system}

\begin{figure}[t]
  \centering
  \includegraphics[width=0.7\linewidth]{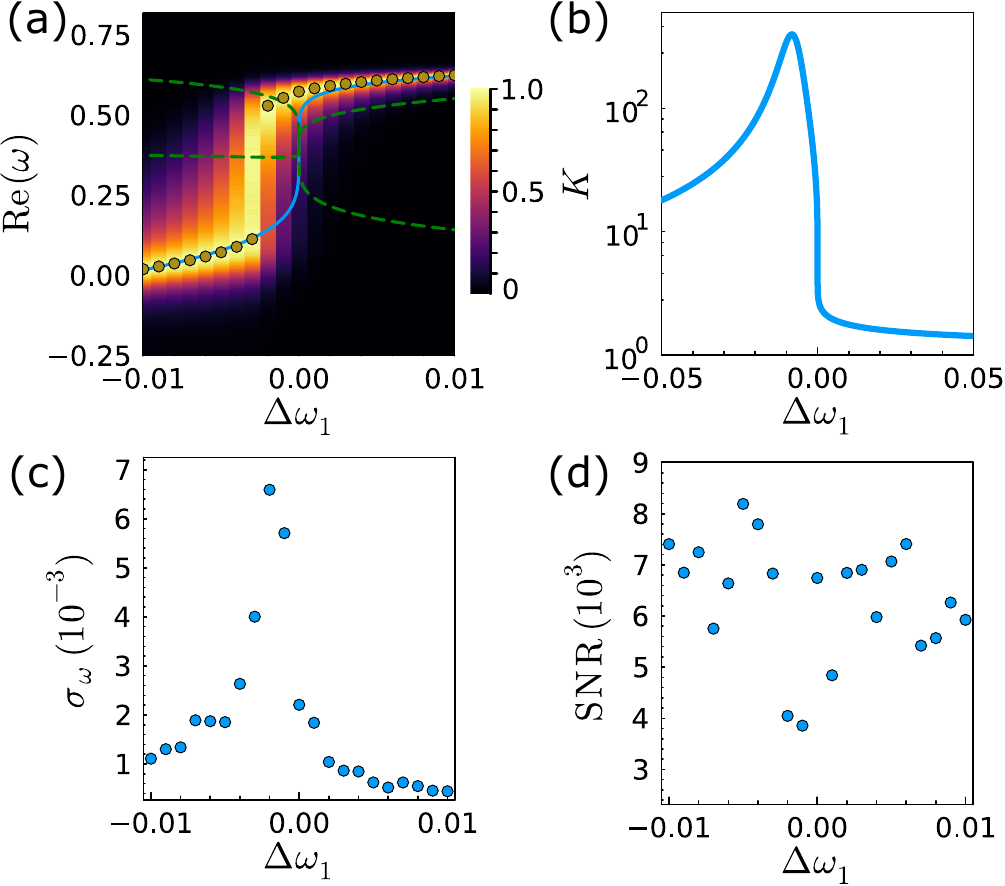}
  \caption{Nonlinear EP-sensing in three-cavity system. (a) Power spectrum plotted against detuning $\Delta\omega_1$ obtained from numerical simulations. The solid blue line denotes the steady-state frequency of the noise-free system. The green dashed lines denote the doubly degenerate unstable steady-state frequency of the noise-free system. The brown dots represent the peak frequencies obtained from curve fitting. (b) Petermann factor versus detuning $\Delta\omega_1$ in the noise-free case. (c) Uncertainties of the peak frequency extracted from the power spectrum. (d) SNRs of the three-cavity-sensing scheme. $\gamma_1=2$, $\Gamma=0.01$, $D=0.3$.}
    \label{fig:threecavity}
\end{figure}

In the main text, we have investigated a nonlinear-EP sensing scheme using an exemplary two-cavity system.  In this section, we investigate a three-cavity scheme similar to that studied in Ref.~\cite{bai2023nonlinear}.  This system corresponds to the case where the steady-state frequency features an EP, but the corresponding instantaneous Hamiltonian is not near an EP.

The Langevin equations for the field $\bm{\alpha}=[\alpha_1,\alpha_2,\alpha_3]^T$ are
\begin{align}
  i\dot{\bm{\alpha}}=\bm{H}\bm{\alpha}+\bm{\xi},
\end{align}
where 
\begin{align}
  \bm{H} = \left[\begin{array}{ccc}
      \omega_1+i\gamma_1+i\Gamma(1-|\alpha_1|^2) & g_1 & 0 \\
      g_1 & \omega_2-i\gamma_2 & g_2 \\
      0 & g_2 & \omega_3+i\gamma_3
  \end{array}\right],
\end{align}
with the noise term $\bm{\xi}=[\xi_1,\xi_2,\xi_3]^T$ satisfying $\langle\xi_i^{\ast}(t)\xi_j(t^{\prime})\rangle=D_{ij}\delta(t-t^{\prime})$. For simplicity, we assume that $D_{ij}=D\delta_{ij}$. In the absence of noise, the steady-state frequencies of the system satisfy the equation $\det{(\bm{H}-\omega\bm{I})}=0$. By requiring the real and imaginary part of the determinant to be zero, we find the steady-state frequencies satisfy a 5th-order polynomial equation.
\begin{align}
    \sum_{i=0}^5a_i\omega^i=0.
\end{align}
By setting $\sum_{i=0}^5a_i\omega^i=(\omega-\omega_0)^5$ with a preselected $\omega_0$, we can solve for the parameters required to achieve a 5th-order EP. Here We choose $\omega_1=0,\omega_2=0.258,\omega_3=0.742,g_1=1,g_2=0.456,\gamma_2=1.275,\gamma_3=0.087$. 

We first consider the noise-free case. In Fig.~\ref{fig:threecavity}(a), the solid blue line shows the stable steady-state frequency versus $\Delta\omega_1$, the green dashed lines show the degenerate unstable steady-state frequency, with all five roots coincide at $\Delta\omega_1=0$. Near the EP, the frequency scales as $\omega\propto\Delta\omega_1^{1/5}$, leading to divergent susceptibility $|\partial\omega/\partial\Delta\omega_1|\propto\Delta\omega_1^{-4/5}$. The instantaneous Hamiltonian, however, is not near an EP, resulting in a finite Petermann factor $K$, as shown in Fig.~\ref{fig:threecavity}(b). Hence, the SNR is divergent at the EP in the noise-free scenario. 

In the presence of noise, we observe a peak frequency splitting and a degradation of the susceptibility near the original EP, as show by the heatmap of Fig.~\ref{fig:threecavity}(a). The brown dots represent the peak frequencies extracted from the power spectrum. Fig.~\ref{fig:threecavity}(c) shows the uncertainties of the dominant peak frequency, which are enhanced near the value of $\Delta\omega_1$ where the peak frequency splitting occurs. As for the SNR, we do not observe any clear improvement near the original EP, as shown in Fig.~\ref{fig:threecavity}(d).

\section{Analysis of the quantum limit}

\begin{figure}[t]
    \centering
    \includegraphics[width=0.7\linewidth]{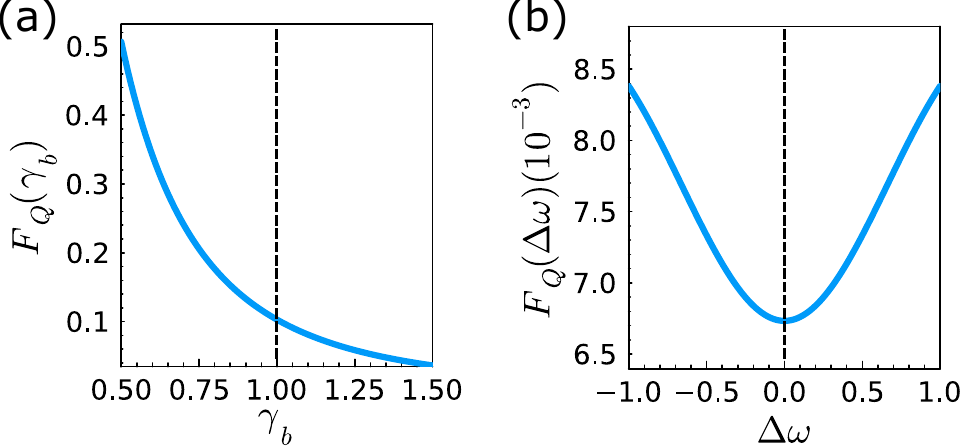}
    \caption{QFI of steady state. (a) QFI versus loss rate $\gamma_b$. $\Delta\omega=0$. (b) QFI versus detuning $\Delta\omega$. $\gamma_b=1$. The dashed lines indicate the location of classical noise-free EP. In (a) and (b), the other model parameters used are $\gamma_a=2$, $\Gamma=10$, $g=0.5$.}
    \label{fig:QFI}
\end{figure}

For $\Gamma\gg\gamma_a$, we have $\langle\hat{a}^{\dagger}\hat{a}\rangle$, $\langle\hat{b}^{\dagger}\hat{b}\rangle\sim 1$. In this quantum limit, the system dynamics are described by the quantum master equation \eqref{eq:mastereq}. In quantum metrology, a widely used quantity is the quantum Fisher information (QFI). Given a state $\rho$, the achievable precision in the statistical estimation of the parameter $\theta$ is constrained by the QFI of the state $F_{Q}(\theta)$ through the quantum Cram\'er-Rao bound \cite{cramer1999mathematical}
\begin{align}
    (\Delta\theta)^2 \ge \frac{1}{mF_Q(\theta)},
\end{align}
where $m$ is the number of independent experimental repetitions. A larger QFI means higher achievable precision. For a density matrix $\rho(\theta)$, we have the spectral decomposition $\rho(\theta)=\sum_i\lambda_i|\lambda_i\rangle\langle\lambda_i|$ with $\lambda_i$ and $|\lambda_i\rangle$ being the eigenvalues and eigenvectors, respectively. The QFI is given by \cite{braunstein1994statistical,paris2009quantum}
\begin{align}
    F_Q(\theta)=2\sum_{i,j}\frac{|\langle\lambda_i|\partial_{\theta}\rho|\lambda_j\rangle|^2}{\lambda_i+\lambda_j},
\end{align}
where the sums include only terms with $\lambda_i+\lambda_j\neq 0$. Hence, the maximal precision is determined by the change of state $\rho(\theta)$ in response to the parameter $\theta$. 

We focus on the steady state $\rho_{\text{st}}$ under the change of the loss rate $\gamma_b$ or the detuning $\Delta\omega$. To solve the quantum master equation \eqref{eq:mastereq}, we truncate the Fock space at $N=10$ for operators $\hat{a}$ and $\hat{b}$. The resulting QFI of steady state $\rho_{\text{st}}$ is depicted in Fig.~(\ref{fig:QFI}), where we observe no QFI divergence near the classical noise-free EP. This indicates there are no sensing enhancements in steady-state measurement, which is consistent with our analysis of classical regime. This analysis is limited to the steady state; exploring whether the system's temporal dynamics could reveal further information is a subject for future study.

\section{Potential experimental platforms}

Several experimental platforms, which have previously been employed to study EPs and related topics, could be used to validate our present results on nonlinear EPs:

\begin{itemize}
\item \textbf{Microring resonators:} Arrays of coupled microring resonators fabricated from multilayer quantum wells have been used to realize lasing of 1D and 2D topological edge modes \cite{partoEdgeModeLasing1D2018,zhaoTopologicalHybridSilicon2018,bandresTopologicalInsulatorLaser2018}. In such systems, saturable gain can be introduced through the quantum wells, while coupling and loss can be tuned via the inter-ring spacing and chromium coatings, respectively.

\item \textbf{Micropillars:} Similar to microring resonators, micropillars have been used to realize lasing with topological edge modes \cite{st-jeanLasingTopologicalEdge2017}. These resonators consist of quantum wells positioned between two distributed Bragg reflectors.

\item \textbf{Whispering-gallery-mode resonators:} These systems have been widely used to explore EP-related phenomena, such as Refs.~\cite{pengLossinducedSuppressionRevival2014,chen2017exceptional}. In these systems, saturable gain can be introduced via Raman or Brillouin lasing processes, while loss can be tuned by a metal-coated tip, and the coupling can be tuned by adjusting the resonator spacing.
\end{itemize}\clearpage
\end{widetext}

\end{document}